\nonstopmode

\newcommand{\Cal}[1]{{\cal #1}}
\newcommand{\ie}{i.e.{}}
\newcommand{\eg}{e.g.{}}
\newcommand{\etal}{\textit{et al.}}
\newcommand{\U}[1]{\,{\rm{#1}}}
\newcommand{\I}[1]{_{\textrm{\scriptsize #1}}}
\newcommand{\imag}{{\rm i}}
\newcommand{\euler}{\textrm{e}}
\newcommand\laplace{\Delta}
\newcommand{\mat}[1]{\hbox{\boldmath{$#1$}\unboldmath}}
\newcommand{\Sum}{\sum\limits}
\newcommand{\Lim}{\lim\limits}
\newcommand{\transpose}{{}^{\textrm{\scriptsize T}}}
\newcommand{\unitmatrix}{\mat{\mathbbm{1}}}
\newcommand{\differential}{\>\textrm{d}}
\newcommand{\bra}[1]{\left<\right.\!#1\!\left.\right|}
\newcommand{\ket}[1]{\left|\right.\!#1\!\left.\right>}
\newcommand{\eV}{\U{eV}}
\newcommand{\angstrom}{\U{\hbox{\AA}}}
\newcommand{\HF}{(HF)$_{\infty}$}

\documentclass[prb,showpacs,nopreprintnumbers,twocolumn]{revtex4}
\usepackage{graphicx,subeqnarray,bbm,bm}
\usepackage[nativepdf,bookmarks,bookmarksopen,bookmarksnumbered,
raiselinks,%
pdftitle={Ab initio Green's function formalism for band structures},%
pdfauthor={Christian Buth, Uwe Birkenheuer, Martin Albrecht, Peter Fulde},%
pdfsubject={Solid State Physics},%
pdfkeywords={ab initio, Green's function, ADC, CO-ADC, algebraic %
diagrammatic construction, crystal orbital, Wannier function,%
band structures}]{hyperref}

\begin{document}
\title{\emph{Ab initio} Green's function formalism for band structures}
\date{20 November 2005}
\author{Christian Buth}
\email[Corresponding author. Electronic address: ]{Christian.Buth@web.de}
\author{Uwe Birkenheuer}
\author{Martin Albrecht}
\altaffiliation{Present address: Theoretische Chemie, Universit\"at Siegen,
57068~Siegen, Germany}
\author{Peter Fulde}
\affiliation{Max-Planck-Institut f\"ur Physik komplexer Systeme, N\"othnitzer
Stra\ss{}e~38, 01187~Dresden, Germany}

\begin{abstract}
Using the Green's function formalism, an \emph{ab initio}
theory for band structures of crystals is derived
starting from the Hartree-Fock approximation. It is
based on the algebraic diagrammatic construction scheme for
the self-energy which is formulated for
crystal orbitals~(CO-ADC). In this approach,
the poles of the Green's function are determined by solving a
suitable Hermitian eigenvalue problem. The method is not only applicable
to the outer valence and conduction bands, it is also stable for inner
valence bands where strong electron correlations are effective. The key to
the proposed scheme is to evaluate the self-energy in terms of Wannier
orbitals before transforming it to a crystal momentum representation.
Exploiting the fact that electron correlations are mainly local, one can truncate
the lattice summations by an appropriate configuration selection scheme.
This yields a flat configuration space; \ie, its size scales only linearly
with the number of atoms per unit cell for large systems and, under certain
conditions, the computational effort to determine band
structures also scales linearly. As a first application of
the new formalism, a lithium fluoride crystal has been chosen.
A minimal basis set description is studied, and
a satisfactory agreement with previous theoretical
and experimental results for the fundamental band gap
and the width of the F$\,2p$~valence band complex is obtained.
\end{abstract}

%
%
%

\preprint{arXiv:cond-mat/0409078}
\maketitle

\section{Introduction}

Band structures are of fundamental interest to the solid-state physicist as they
reveal important properties of crystals.~\cite{Ashcroft:SSP-76,Pisani:HF-88,%
Callaway:QT-91,Fulde:EC-95,Pisani:QM-96,Fulde:WF-02} The determination of band
structures on \emph{ab initio} level is therefore an important issue in
theoretical solid state physics. Nowadays predominantly density functional
theory~(DFT)~\cite{Hohenberg:IEG-64,Kohn:SC-65} with the local density
approximation~(LDA) or its variants are employed to treat this
problem.~\cite{Callaway:QT-91,Fulde:EC-95} DFT focuses on the
ground-state electron density and yields ground-state properties.
In several cases,
the interpretation of the Kohn-Sham orbital energies as quasiparticle
energies also turns out to be successful and particular attention has
been paid to the~LDA, due to its numerical simplicity. However, in
insulating materials, the LDA tends to underestimate the band gap,
\eg, Ref.~\onlinecite{Ching:BT-95}.

As is well known, it is difficult to improve band structure
calculations based on density functional theory.
This is in particular the case when electron correlations
are strong such that in quite some
cases~\cite{Lowdin:QT-56,Lowdin:BT-62,Ashcroft:SSP-76,Cederbaum:CE-86,%
Buth:IO-03,Buth:IM-03}
the one-electron picture becomes inappropriate and corrections are needed.
Nevertheless, a number of ingenious methods have been devised to improve
band structure results based on density functional theory.
For example, the calculations have been supplemented by a
GW~treatment, \eg, Refs~\onlinecite{Hedin:NM-65,Hott:GW-91,Fulde:EC-95}, or
the so-called LDA+$U$ method.~\cite{Anisimov:FP-97} Other
improvements concern the use
of optimized effective potentials~\cite{Grabo:SC-98} such as the
exact exchange potential~\cite{Stadele:EE-99} instead of its local,
simplified form. Time-dependent DFT is another route
to treat excited states more rigorously.
Furthermore, LDA calculations have been coupled with various
extensions of the coherent potential approximation~(CPA)
like the dynamical CPA~\cite{Kakehashi:MC-92,Kakehashi:DC-02}
or dynamical mean field
theory~(DMFT).~\cite{Metzner:CL-89,Metzner:ECL-89,
Muller-Hartmann:CF-89,Pruschke:AN-95,Georges:DM-96,
Jarrell:SC-01} Also Faddeev's method~\cite{Faddeev:-61}
of treating the three-particle $t$-matrix has been applied in
conjunction with LDA band structure calculations.~\cite{Unger:EE-94}
The same holds true for different forms of the projection
operator technique.~\cite{Fulde:EC-95}
Yet one must face the fact that the various approximations
remain uncontrolled and therefore have to be reconsidered
from case to case.

When one is aiming at controlled approximations, two different
routes offer themselves. One is bound to quantum Monte
Carlo~(QMC) calculations, \eg, Ref.~\onlinecite{Foulkes:QMC-01}.
This is a simple and straightforward way of dealing with
the many-body problem. Yet it suffers from one serious
shortcoming, the so-called sign problem when treating fermions.
Therefore, we are pursuing the second route. It starts from
Hartree-Fock band structures
and improves them by including correlations by means of
many-particle techniques similar to the ones employed for
molecules.~\cite{Ladik:QT-88,Grafenstein:VB-93,%
Fulde:EC-95,Grafenstein:VB-97,Albrecht:LO-98,Ladik:PS-99,%
Albrecht:AA-00,Albrecht:LO-01,Albrecht:LOB-02,Albrecht:LAI-02,%
Fulde:WF-02,Bezugly:MC-04,Buth:MC-05}
Thereby one can take advantage of program packages like
\textsc{wannier} which has been written by
Shukla~\etal~\cite{Shukla:EC-96,Shukla:WF-98}
or \textsc{crystal} which originates from
Torino.~\cite{Pisani:HF-88,Pisani:QM-96} Both provide Hartree-Fock
bands as well as Wannier orbitals for crystals. These
orbitals are especially suited for an \emph{ab initio} treatment
of electron correlations because of their local
character.~\cite{Fulde:EC-95,Fulde:WF-02}

A number of correlation calculations have been
performed.~\cite{Fulde:EC-95,Fulde:WF-02,Ladik:QT-88,Ladik:PS-99}
A local Hamiltonian approach~\cite{Grafenstein:VB-93,%
Grafenstein:VB-97,Albrecht:LO-98,Albrecht:AA-00,%
Albrecht:LAI-02,Bezugly:MC-04} was shown to
improve Hartree-Fock energy bands substantially.
Suhai used Toyozawa's electronic polaron model~\cite{Toyozawa:EP-54}
to repartition MP2 pair energies to estimate quasiparticle band
structures.~\cite{Suhai:QP-83,Ladik:QT-88,Sun:SO-96,Ladik:PS-99,%
Hirata:MB-00,Pino:LT-04}
By inserting the orbital energies into the energy-dependent
self-energy, this model was shown to be a special case of
the outer valence Green's functions~(OVGFs)~\cite{Cederbaum:OB-75,%
Cederbaum:TA-77,Niessen:CM-84} that were derived in terms of
crystal orbitals by
Liegener.~\cite{Liegener:TO-85,Ladik:QT-88,Ladik:PS-99,Pino:LT-04}
By applying the approximation to the self-energy of
Igarashi~\etal,~\cite{Igarashi:LA-94,Albrecht:LO-01}
quasiparticle band structures were obtained
by Albrecht~\etal.~\cite{Albrecht:LO-01,Albrecht:LOB-02,%
Albrecht:LAI-02}

In the present paper, we derive the algebraic diagrammatic
construction~(ADC) scheme~\cite{Schirmer:PP-82,Schirmer:GF-83,%
Cederbaum:GF-98} for crystals. The ADC scheme has proven to be superior to the
OVGF method in molecular studies,~\cite{Niessen:CM-84} and
numerous works have been carried out over the last two decades, \eg,
Refs.~\onlinecite{Cederbaum:CE-86,Buth:IO-03,Buth:IM-03},
including studies of oligomers and clusters chosen
to model infinite chains or crystals.~\cite{Deleuze:SB-96,%
Golod:VC-99,Santra:ED-01}
The ADC scheme is a method to approximate
the Feynman-Dyson perturbation series for the
self-energy and contains sums of certain proper and improper
diagrams to infinite order.~\cite{Schirmer:GF-83,Schirmer:SE-89,Santra:CAP-02}
The basic properties of the
ADC scheme and their derivations, as applied to molecules,
are found in Refs.~\onlinecite{Schirmer:CF-91,Schirmer:SC-96,
Mertins:AP1-96}. Among those size-extensivity is an important
one since it is crucial when solids are considered. Furthermore,
the ADC method is known to be robust and facilitates also
to study strong electron correlations due to the
efficient and stable evaluation of the one-particle Green's function
in terms of a Hermitian eigenvalue problem. In molecules
strong correlations for instance occur when inner valence electrons are
treated.~\cite{Cederbaum:CE-86,Buth:IO-03,Buth:IM-03}
With the crystal orbital ADC (CO-ADC) devised here, the spectral
representation of the one-particle Green's function
is obtained.
We suggest a formulation of the expressions
in Wannier representation instead of crystal momentum representation
in order to be able to exploit the local character of the correlation
hole around an electron especially in the case of crystals with
a large unit cell.

As a first application of the CO-ADC method to a three-dimensional
crystalline solid, we have chosen lithium fluoride
which occurs in nature as the mineral griceite.
LiF not only has a wide range of technological
applications like in x-ray monochromators or
in filters for ultraviolet radiation,
\eg, Ref.~\onlinecite{Shirley:DT-96} (and References.{}
therein), but is also interesting for a number of
fundamental physical reasons. It is
considered to be the ``most ionic substance''
and a prototypical insulator which
manifests in its very large
fundamental band gap of$\,$%
\footnote{The experimental band gap of LiF
is communicated in the early work of Roessler
and Walker to be~$13.60 \pm 0.06 \eV$.~\cite{Roessler:ES-67,%
Poole:EB-75,Poole:ES-75}
Piacentini~\cite{Piacentini:NI-75}
estimates the band gap of LiF to be~$14.5 \eV$.
This value is refined in Ref.~\onlinecite{Piacentini:TR-76}
to~$14.2 \pm 0.2 \eV$.
Shirley~\etal~\cite{Shirley:DT-96}
communicate~$14.1 \pm 0.1 \eV$ from the
dissertation of Himpsel.}
$14.1 \pm 0.1 \eV$ that is the largest one found in nature apart
from exotic systems. Some authors even consider~LiF to be
comparable to a He--Ne rare-gas
solid.~\cite{Shirley:DT-96}
Its optical spectra show strong excitonic
effects, which complicates the experimental
determination of the band gap~\cite{Piacentini:TR-76}
by optical spectroscopy.
Similarly, many-particle effects have to be accounted
for in the measurement of the widths
of the F$\,2p$ and F$\,2s$~valence bands.

LiF has been thoroughly studied both
experimentally
and theoretically.~\cite{Wyckoff:CS-68,Poole:EB-75,%
Poole:ES-75,Kunz:SE-82,%
Prencipe:AI-95,Shirley:DT-96,Doll:CP-97,%
Shukla:WF-98,
Wang:QP-03,Albrecht:LOB-02,Albrecht:LAI-02}
Poole~\etal{} survey early
experimental and theoretical
data.~\cite{Poole:EB-75,Poole:ES-75}
Recent studies of LiF comprise
density functional theory calculations
at the LDA level supplemented by
an inclusion of correlation
effects in terms of the
GW~approximation.~\cite{Hedin:NM-65,Shirley:DT-96,Wang:QP-03}
\emph{Ab initio} investigations comprise
Hartree-Fock studies of ground-state
properties~\cite{Prencipe:AI-95,Shukla:WF-98}
and of the band structure.~\cite{Kunz:SE-82,%
Albrecht:LOB-02,Albrecht:LAI-02}
An accurate treatment of electron correlations for
the ground-state properties of~LiF (and other alkali halides)
has been carried out by Doll and
Stoll.~\cite{Doll:CP-97}
Quasiparticle band structures have been obtained
by Kunz~\cite{Kunz:SE-82} on the basis of the
electronic polaron model~\cite{Toyozawa:EP-54} and
by Albrecht~\cite{Albrecht:LOB-02}
with the Green's function approach of
Igarashi~\etal.~\cite{Igarashi:LA-94}

The article is structured as follows: Section~\ref{sec:band} puts forward
an \emph{ab initio} foundation for band structures based on the
one-particle Green's function. In Secs.~\ref{sec:dynamic} and
\ref{sec:static}, we derive the CO-ADC approximation to the
self-energy in terms of Wannier orbitals. Section~\ref{sec:poledyson}
combines the results of the previous sections to formulate the
band structure problem as a Hermitian matrix eigenvalue problem.
Criteria for the truncation of lattice sums are discussed in
Sec.~\ref{sec:onion}. In Sec.~\ref{sec:LiF_crystal},
the application of CO-ADC theory to lithium fluoride
is described.
Conclusions are drawn in Sec.~\ref{sec:conclusion}.

\section{Band Structures}
\label{sec:band}

The nonrelativistic Hamiltonian of a crystal in atomic
units~\cite{Ladik:QT-88,Fulde:EC-95,Ladik:PS-99} reads
\begin{equation}
  \label{eq:Operator_Hamilton}
  \begin{array}{rcl}
    \hat H &=& \Sum_{n=1}^N \Bigl[ - \frac{1}{2} \laplace_n
      - \Sum_{i=1}^{N_0} \Sum_{A=1}^M \frac{Z_{\vec R_i A}}
      {|\vec r_n - \vec r_{\vec R_i A}|} \Bigr ] \\
    &&{} + \frac{1}{2} \, \Sum_{\scriptstyle m,n=1 \atop \scriptstyle m \neq n}^N \frac{1}{|\vec r_m -
      \vec r_n|} \\
    &&{} \stackrel{\textstyle + \frac{1}{2} \, \Sum_{i,j=1}^{N_0} \Sum_{A,B=1}^M
      \frac{Z_{\vec R_i A} Z_{\vec R_j B}}{|\vec r_{\vec R_i A} -
      \vec r_{\vec R_j B}|}} {\qquad\scriptstyle i \neq j \lor A \neq B \hfill} \; ,
  \end{array}
\end{equation}
where fixed nuclei are assumed. Here $N$~is the number
of electrons in the crystal, $N_0$~denotes the number of
unit cells, and $M$~is the number of nuclei
per unit cell. $Z_{\vec R_i A} \equiv Z_A$~stands for the charge of
nucleus~$A$ in unit cell~$\vec R_i$, and $|\vec r_n -
\vec r_{\vec R_j A}|$~is the distance between the $n$-th
electron and the $A$-th nucleus in unit cell~$\vec R_j$.
Finally, $|\vec r_m - \vec r_n|$~represents the distance between the
$m$-th and $n$-th electron and $|\vec r_{\vec R_i A}
- \vec r_{\vec R_j B}|$~denotes the distance between the nuclei~$A$,
$B$ of charge~$Z_{\vec R_i A} \equiv Z_A$, $Z_{\vec R_i B} \equiv Z_B$
in unit cells~$\vec R_i$, $\vec R_j$.

The Schr\"odinger equation with the Hamiltonian~(\ref{eq:Operator_Hamilton})
can be solved for the ground state in a restricted, closed-shell Hartree-Fock
approximation,~\cite{Lowdin:QT-56,Lowdin:BT-62,Re:SC-67,Andre:EH-67,Andre:ET-67}
yielding a set of spin Bloch orbitals~$\psi_{\vec k \, p}(\vec r)$
where $\vec k$~is the crystal momentum and $p$~is a band
index.~\cite{Lowdin:QT-56,Lowdin:BT-62,Re:SC-67,Andre:EH-67,%
Andre:ET-67,Pisani:HF-88,Ashcroft:SSP-76,Callaway:QT-91,%
Fulde:EC-95,Pisani:QM-96} We use these
orbitals to represent the Hamiltonian~(\ref{eq:Operator_Hamilton})
in second quantization, yielding the so-called
Bloch or crystal momentum representation.~\cite{Callaway:QT-91}
It is decomposed according to M\o{}ller and
Plesset~\cite{Szabo:MQC-89,McWeeny:MQM-92,Helgaker:MES-00}
into the Hartree-Fock part~$\hat H\I{HF}^{\rm BF}$ and the
residual interaction~$\hat H\I{res}^{\rm BF}$, a suitable form to
apply perturbation theory:
\renewcommand{\arraystretch}{1.5}%
\begin{equation}
  \label{eq:Bloch_Hamilton}
  \begin{array}{rcl}
    \hat H &=& \hat H\I{HF}^{\rm BF} + \hat H\I{res}^{\rm BF} \; , \\
    \hat H\I{HF}^{\rm BF} &=& \Sum_{\vec k \, p} \varepsilon_{\vec k \, p} \,
      \hat c_{\vec k \, p}^{\dagger} \, \hat c_{\vec k \, p} \; , \\
    \hat H\I{res}^{\rm BF} &=& \Sum_{\vec k_1 \, p, \  \vec k_2 \, q} W_{\vec
      k_1 \, p \; \vec k_2 \, q} \> \hat c_{\vec k_1 \, p}^{\dagger} \,
      \hat c_{\vec k_2 \, q} \\
    &&{} + \frac{1}{2} \Sum_{\scriptstyle \vec k_1 \, p, \ \vec k_2 \, q, \atop
      \scriptstyle \vec k_3 \, r,
      \  \vec k_4 \, s} V_{\vec k_1 \, p \; \vec k_2 \, q \; \vec k_3 \, r \; \vec
      k_4 \, s} \\
    &&{} \times \hat c_{\vec k_1 \, p}^{\dagger} \, \hat c_{\vec k_2 \, q}^{\dagger}
      \, \hat c_{\vec k_4 \, s} \, \hat c_{\vec k_3 \, r} \; .
  \end{array}
\end{equation}
\renewcommand{\arraystretch}{1}%
The energy bands (or Bloch orbital energies)
are denoted by~$\varepsilon_{\vec k \, p}$,
and the operators~$\hat c_{\vec k \, p}^{\dagger}$ ($\hat c_{\vec k \, p}$)
create (annihilate) electrons in~$\psi_{\vec k \, p}(\vec r)$.
The negative of the Hartree-Fock potential is given by~$W_{\vec k_1 \, p \; \vec k_2 \, q}
= - \Sum_{\vec k \, i} V_{\vec k_1 \, p \; \vec k \, i \; [\vec k_2 \, q
\; \vec k \, i]} \> n_{\vec k \, i}$,
the occupation numbers~$n_{\vec k \, p}$ being unity for occupied and zero
for unoccupied (or virtual) Bloch orbitals (and $\bar n_{\vec k \, p} =
1 - n_{\vec k \, p}$). Generally, we use the band indices~$i$,
$j$, $m$, $n$, \ldots{} to denote occupied Hartree-Fock bands,
$a$, $b$, $c$, $d$, \ldots{} for unoccupied bands, and
$p$, $q$, $r$, $s$, \ldots{} for bands which are occupied or unoccupied.
The two-electron integrals in Eq.~(\ref{eq:Bloch_Hamilton}) are defined with
respect to Bloch orbitals by~\cite{Szabo:MQC-89,McWeeny:MQM-92,Helgaker:MES-00}
\begin{equation}
  \label{eq:twoint}
  \begin{array}{rcl}
    V_{\vec k_1 \, p \; \vec k_2 \, q \; \vec k_3 \, r \; \vec
      k_4 \, s} &\!=\!& \int \!\! \int \psi_{\vec k_1 \, p}^\dagger(\vec r_1) \,
       \psi_{\vec k_2 \, q}^\dagger(\vec r_2) \> \frac{1}{|\vec r_1 - \vec r_2|}
       \qquad\ \ \ \nonumber \\
    &&{} \times \psi_{\vec k_3 \, r}(\vec r_1) \, \psi_{\vec k_4 \, s}(\vec r_2)
      \, \differential^3 r_1 \differential^3 r_2 \; .
  \end{array}
\end{equation}
The antisymmetrized two-electron integrals are
\begin{equation}
  \label{eq:antitwoint}
    V_{\vec k_1 \, p \; \vec k_2 \, q \; [\vec k_3 \, r \; \vec
      k_4 \, s]} = V_{\vec k_1 \, p \; \vec k_2 \, q \; \vec k_3 \, r \; \vec
      k_4 \, s} - V_{\vec k_1 \, p \; \vec k_2 \, q \; \vec k_4 \, s \; \vec
      k_3 \, r}
\end{equation}
and consist of a Coulomb term~$V_{\vec k_1 \, p \; \vec k_2 \, q \; \vec k_3 \, r
\; \vec k_4 \, s}$ and an exchange term~$V_{\vec k_1 \, p \; \vec k_2 \, q \;
\vec k_4 \, s \; \vec k_3 \, r}$.
All crystal momentum vectors in Eqs.~(\ref{eq:Bloch_Hamilton}),
(\ref{eq:twoint}), and (\ref{eq:antitwoint}) are restricted to
the first Brillouin zone. Furthermore, we assume that the first Brillouin
zone is discretized by the Born von K\'arm\'an boundary conditions, or
equivalently, that numerical integrations in reciprocal space are carried
out in terms of a Monkhorst-Pack net.~\cite{Monkhorst:SP-76,Pisani:HF-88,%
Pisani:QM-96} In both cases, integrations over the Brillouin zone are
approximated by finite sums.

The one-particle Green's function (or particle propagator) in terms of Bloch
orbitals~\cite{Schirmer:GF-83,Mattuck:FD-76,Fetter:MP-71,Gross:MP-91,
Callaway:QT-91} reads
\begin{equation}
  \label{eq:k-prop}
  G_{\vec k \, p \; \vec k^{\prime} \, q}(t, t^{\prime})
   = (-\imag) \, \bra{\Psi_0^N} \hat T[\hat c_{\vec k p}(t) \,
    \hat c_{\vec k^{\prime} q}^{\dagger}(t^{\prime})] \ket{\Psi_0^N} \; ,
\end{equation}
with Wick's time-ordering operator~$\hat T$. $\ket{\Psi_0^N}$~is the exact
ground state of the $N$~particle system. Due to translational symmetry of the
Hamiltonian~(\ref{eq:Operator_Hamilton}), the one-particle Green's function
depends only on one crystal momentum,~\cite{Callaway:QT-91}
\ie, $G_{\vec k \, p \; \vec k^{\prime} \, q}(t, t^{\prime}) =
\delta_{\vec k , \vec k^{\prime}} \, G_{pq}(\vec k,t, t^{\prime})$.

Similarly, the time independence of the Hamiltonian~(\ref{eq:Operator_Hamilton})
causes the particle propagator~(\ref{eq:k-prop}) to depend only on the time difference~$t -
t^{\prime}$.~\cite{Mattuck:FD-76,Fetter:MP-71,Gross:MP-91}
Fourier transforming~$G_{pq}(\vec k,t - t^{\prime})$ with respect to~$t -
t^{\prime}$ yields the one-particle Green's function in energy
space~$G_{pq}(\vec k,\omega)$ which can be recast in terms of
the spectral or Lehmann representation as~\cite{Mattuck:FD-76,%
Fetter:MP-71,Gross:MP-91}
\begin{subeqnarray}
  \label{eq:p-spectrum}
  G_{pq}(\vec k,\omega) &=& \Sum_{n \in \{N+1\}}
  \slabel{eq:p-spectral}
  \frac{y_p^{(n)}(\vec k) \; y_q^{(n)*}(\vec k)}
       {\omega + A_n(\vec k) + \imag \eta} \\
  &&{} + \Sum_{n \in \{N-1\}} \nonumber
  \frac{x_p^{(n)}(\vec k) \; x_q^{(n)*}(\vec k)}
       {\omega  + I_n(\vec k) - \imag \eta} \\
  \slabel{eq:p-decomp}
  &=& G_{pq}^+(\vec k,\omega) + G_{pq}^-(\vec k,\omega) \; .
\end{subeqnarray}
Here $n$~numerates the excited states of the $N\pm 1$~particle system.
The negative of the pole positions in Eq.~(\ref{eq:p-spectral}) is given
by either the electron affinities~$A_n(\vec k) = E_0^N - E_n^{N+1}(\vec k)$
or the ionization potentials $I_n(\vec k) = E_n^{N-1}(-\vec k) - E_0^N$
where $E_n^{N \pm 1}(\pm\vec k)$~are the energies of the different
excited states of the $N \pm 1$~particle system.
The summands~$\pm \imag \eta$ are required to ensure the convergence of
the Fourier transformation. Whenever such a factor occurs,
we implicitly take the limit~$\eta \to 0^+$.~\cite{Mattuck:FD-76,%
Fetter:MP-71,Gross:MP-91}

The pole strengths in Eq.~(\ref{eq:p-spectral})
are given by the transition amplitudes~\cite{Weikert:BL-96}
\begin{equation}
  \begin{array}{rcl}
    \label{eq:transamp}
    y_p^{(n)}(\vec k) &=& \bra{\Psi_0^N}
      \hat c_{\vec k \, p} \ket{\Psi_n^{N+1}(\vec k)} \; , \\
    x_p^{(n)}(\vec k) &=& \bra{\Psi_n^{N-1}(-\vec k)}
      \hat c_{\vec k \, p} \ket{\Psi_0^N} \; ,
  \end{array}
\end{equation}
where $\ket{\Psi_n^{N \pm 1}(\pm \vec k)}$~denote excited states of the
$N \pm 1$~particle system with crystal momenta~$\pm \vec k$.
The pole strengths can be interpreted in terms of the spectral intensities
observed in photoelectron spectroscopy experiments, similarly to the
molecular case discussed in Ref.~\onlinecite{Cederbaum:TA-77}.

To obtain electron affinities and ionization potentials from the band
structure of three-dimensional crystals, one has to specify the energy
of the added (removed) electron at the Fermi level of
the neutral crystal.~\cite{Pisani:UH-96,Birkenheuer:EC-97}
This is usually done by introducing chemical
potentials~$\mu^{\pm}$ which are added to the pole positions of
the one-particle Green's function, \ie, $A_n(\vec k) + \mu^+$ and
$I_n(\vec k) + \mu^-$.

The (fundamental) band gap
is the smallest difference between the energies
for removing an electron from and attaching an
electron to an $N$~particle system:
\begin{equation}
  \label{eq:bandgap}
  \begin{array}{rcl}
    E_{\rm gap} &=& (E^{N-1} - E_0^N) - (E_0^N - E^{N+1}) \\
    &=& I - \mu^- - A + \mu^+ \; .
  \end{array}
\end{equation}
Here $A$~stands for the largest electron affinity
and $I$~designates its smallest ionization
potential. Correspondingly,
$E^{N \pm 1}$~denote the energies of the electron
attachment
(removal) state with
the lowest (highest) energy and $E_0^N$~represents
the ground-state energy.

The self-energy~$\mat \Sigma(\vec k, \omega)$ with respect to
the residual interaction~$\hat H\I{res}^{\rm BF}$
is defined by the Dyson equation~\cite{Fetter:MP-71,Mattuck:FD-76,
Schirmer:GF-83,Gross:MP-91,Callaway:QT-91}
\begin{subeqnarray}
  \label{eq:Dyson}
  \mat G(\vec k, \omega) = \mat G^0(\vec k, \omega) &\!\!+\!\!& \mat G^0(\vec k, \omega)
    \, \mat \Sigma(\vec k, \omega) \, \mat G(\vec k, \omega) \\
  = \mat G^0(\vec k, \omega) &\!\!+\!\!& \mat G^0(\vec k, \omega)
  \slabel{eq:DysonExp}
    \, \mat \Sigma(\vec k, \omega) \, \mat G^0(\vec k, \omega) \quad\  \\
  &\!\!+\!\!& \ldots \; , \nonumber
\end{subeqnarray}
which can be solved formally by
\begin{equation}
  \label{eq:solveDysonFormal}
  \mat G(\vec k, \omega) = [\mat G^0(\vec k, \omega)^{-1}
  - \mat \Sigma(\vec k, \omega)]^{-1} \; ,
\end{equation}
where
\begin{equation}
  \label{eq:freeprop}
  G^0_{pq}(\vec k, \omega) = \delta_{\vec k \, p, \; \vec k \, q} \> \Bigl[
  \frac{\bar n_{\vec k \, p}}{\omega - \varepsilon_{\vec k \, p} + \imag \eta}
  + \frac{n_{\vec k \, p}}{\omega - \varepsilon_{\vec k \, p} - \imag \eta}
  \Bigr]
\end{equation}
is the free Green's function with respect to the M\o{}ller-Plesset
partition~(\ref{eq:Bloch_Hamilton}). In fact, for crystals with a band gap,
the occupation numbers in Eq.~(\ref{eq:freeprop}) are independent
of~$\vec k$, \ie, $n_{\vec k \, p} = n_p$ and $\bar n_{\vec k \, p}
= \bar n_p$.

The self-energy can be decomposed into an $\omega$~independent part,
the static self-energy~$\mat \Sigma^{\infty}(\vec k)$, and
an $\omega$~dependent part, the dynamic self-energy~$\mat M(\vec
k,\omega)$,~\cite{Ethofer:SP-69,Winter:CE-72,Cederbaum:TA-77,Schirmer:GF-83}
\begin{equation}
  \label{eq:SigmaSep}
  \mat \Sigma(\vec k,\omega) = \mat \Sigma^{\infty}(\vec k)
  + \mat M(\vec k,\omega) \; ,
\end{equation}
where $\Lim_{\omega \to \pm \infty} \mat M(\vec k,\omega) = \mat 0$.
The dynamic self-energy is considered first in the ensuing
Sec.~\ref{sec:dynamic} as the static self-energy is determined
using the former one in Sec.~\ref{sec:static}.

\section{Dynamic self-energy}
\label{sec:dynamic}
\subsection{Crystal momentum representation}
\label{sec:crystalM}

The partition~(\ref{eq:Bloch_Hamilton}) of the Hamiltonian has the same form
and meaning as the M\o{}ller-Plesset partition for canonical molecular
orbitals~\cite{Szabo:MQC-89,McWeeny:MQM-92,Helgaker:MES-00}
used frequently in molecular physics. Therefore, all equations derived on
the basis of the molecular M\o{}ller-Plesset partition are in complete
analogy to the equations in the crystalline case. The only difference
between molecules and crystals is the occurrence of composite indices in
the Hamiltonian~(\ref{eq:Bloch_Hamilton}), consisting of a crystal momentum
vector~$\vec k$ and a band index~$p$
which vary
independently. Given an equation in terms of molecular orbitals, it can
be written immediately in terms of Bloch orbitals by replacing all molecular
orbital indices by composite indices. Afterwards one can exploit
(translational) symmetry to simplify the equation.

The dynamic self-energy~$\mat M(\vec k,\omega)$ is represented in terms of
the $2p1h$/$2h1p$-propagator (two-particle-one-hole/two-hole-one-particle-propagator)
and possesses the spectral representation~\cite{Ethofer:SP-69,Winter:CE-72,Schirmer:GF-83}
\begin{eqnarray}
  M_{pq}(\vec k,\omega) &=& \sum_{n \in \{N+1\}} \frac{m_p^{+,(n)}(\vec k)
    \> m_q^{+,(n)*}(\vec k)}{\omega - \omega^+_n(\vec k) + \imag \eta} \nonumber \\
  \label{eq:2h1p-spectral}
  &&+ \sum_{n \in \{N-1\}} \frac{m_p^{-,(n)}(\vec k) \> m_q^{-,(n)*}(\vec k)}
    {\omega - \omega^-_n(\vec k) - \imag \eta} \\
  &=& M_{pq}^+(\vec k,\omega) + M_{pq}^-(\vec k,\omega) \; ,
  \nonumber
\end{eqnarray}
where $\omega_n^{\pm}(\vec k)$~denote the pole positions of the
$2p1h$/$2h1p$-propagator and the $m_p^{\pm,(n)}(\vec k)$~are termed Dyson
amplitudes. The retarded dynamic self-energy~$M_{pq}^+(\vec k,
\omega)$ and the advanced dynamic self-energy~$M_{pq}^-(\vec k,\omega)$
are associated with excitations of the $N \pm 1$~particle system,
respectively. Yet the pole positions~$\omega_n^{\pm}(\vec k)$ do
\emph{not} directly correspond to the energies of physical states
as the (dynamic) self-energy is only defined in conjunction with
the Dyson equation~(\ref{eq:Dyson}).

The algebraic diagrammatic construction scheme is a stable and
efficient method to determine
the spectral representation of the dynamic
self-energy~(\ref{eq:2h1p-spectral}). The construction starts by
making the ansatz
  \begin{equation}
    \label{eq:ADC-form}
    \textstyle
    M_{pq}^{\pm}(\vec k,\omega)
      = \vec U_p^{\pm\dagger}(\vec k) \> (\omega \, \unitmatrix
        - \mat K^{\pm}(\vec k) - \mat C^{\pm}(\vec k))^{-1} \>
        \vec U_q^{\pm}(\vec k) \; ,
  \end{equation}
which is termed general algebraic form or ADC~form.
The vectors~$\vec U_q^{\pm}(\vec k)$ are called modified coupling
amplitudes for the crystal orbital~$\vec k \, q$.
Both $\mat K^{\pm}(\vec k)$ and $\mat C^{\pm}(\vec k)$ are Hermitian matrices,
$\mat C^{\pm}(\vec k)$ being referred to as modified interaction
matrices~\cite{Schirmer:GF-83} while $\mat K^{\pm}(\vec k)$ are assumed
to be diagonal.
Explicit expressions for these matrices are given
below in Eq.~(\ref{eq:k-CO-ADC2}).

The ADC~form~(\ref{eq:ADC-form}) reproduces the analytic structure
of the Feynman-Dyson perturbation
series for the dynamic self-energy as can be seen from its
expansion into a geometric series:
\begin{eqnarray}
  \label{eq:ADCsuminfinity}
  M_{pq}^{\pm}(\vec k,\omega) &=& \vec U_p^{\pm\dagger}(\vec k) \,
    \Sum_{n=0}^{\infty} [ (\omega \, \unitmatrix - \mat
    K^{\pm}(\vec k))^{-1} \, \mat C^{\pm}(\vec k) ]^n \nonumber \\
  \label{eq:ADCanalytic}
  &&{} \times (\omega \, \unitmatrix - \mat K^{\pm}(\vec k))^{-1}
    \, \vec U_q^{\pm}(\vec k) \; .
\end{eqnarray}
The $n$-th order approximation to the ADC~form~(\ref{eq:ADC-form})
is constructed by inserting the perturbation expansions
\begin{equation}
  \begin{array}{rcl}
    \mat U^{\pm}(\vec k) &=& \mat U^{\pm\,(1)}(\vec k) + \mat U^{\pm\,(2)}
      (\vec k) + \ldots \; , \\
    \mat C^{\pm}(\vec k) &=& \mat C^{\pm\,(1)}(\vec k) + \mat C^{\pm\,(2)}
      (\vec k) + \ldots \; ,
  \end{array}
\end{equation}
in terms of the residual interaction~$\hat H^{\rm BF}\I{res}$
into Eq.~(\ref{eq:ADCsuminfinity}), and comparing the resulting
expressions with the diagrammatic expansion of the dynamic
self-energy up to $n$-th order.
This constitutes a scheme which is denoted by~ADC($n$). Note that the analytic
structure imposed by Eq.~(\ref{eq:ADCanalytic}) occasionally requires
one to associate linear combinations of the analytic
expression of several diagrams with particular
terms in the above expansion.

In ADC(2) and ADC(3), each entry of the~$\vec U_q^{\pm}(\vec k)$
and the $\mat K^{\pm}(\vec k) + \mat C^{\pm}(\vec k)$
in Eq.~(\ref{eq:ADC-form})
is characterized by one or two, respectively, arrangements
of two particles and one
hole~($2p1h$) for~$M_{pq}^+(\vec k, \omega)$ and two holes and one
particle~($2h1p$) for~$M_{pq}^-(\vec k,\omega)$. In brief notation these
arrangements are referred to as $2p1h$- and $2h1p$-configurations.
All such configurations that can be formed with the one-particle
basis set underlying a Hartree-Fock calculation constitute the
configuration space in ADC(2) and ADC(3). The configuration
space is enlarged in ADC(4) by $3p2h$- and
$3h2p$-configurations.~\cite{Schirmer:GF-83} In general, every
second order in ADC, the configuration space enlarges
by the next higher excitation class. Restricting our discussion
to $2p1h$- and $2h1p$-configurations, we can rewrite
Eq.~(\ref{eq:ADC-form}) as
\begin{widetext}
  \begin{equation}
    \label{eq:ADC-form-expanded}
    M_{pq}^+(\vec k,\omega) = {\Sum_{\scriptstyle \vec l_1, \vec l_2, \vec
        l_1^{\prime}, \vec l_2^{\prime} \atop \scriptstyle i, a, b, i^{\prime},
        a^{\prime}, b^{\prime}}}^{\hspace{-0.5cm}\prime\hspace{0.4cm}}
        U_{\vec k \, p \; ; \; \vec l_1 + \vec l_2 - \vec k \, i \;
        \vec l_1 \, a \; \vec l_2 \, b}^{+ *} \;
        (\omega \, \unitmatrix - \mat K^+(\vec k)
        - \mat C^+(\vec k))^{-1}_{\scriptstyle \vec l_1 + \vec l_2 - \vec k \, i
        \; \vec l_1 \, a \; \vec l_2 \, b \; ; \atop
        \scriptstyle \vec l_1^{\prime} + \vec l_2^{\prime} - \vec k \, i^{\prime} \;
        \vec l_1^{\prime} \, a^{\prime} \; \vec l_2^{\prime} \, b^{\prime}} \,
        U_{\vec k \, q \; ; \; \vec l_1^{\prime} +
        \vec l_2^{\prime} - \vec k \, i^{\prime} \; \vec l_1^{\prime} \,
        a^{\prime} \; \vec l_2^{\prime} \, b^{\prime}}^+ \; .
  \end{equation}
\end{widetext}
The prime on the summation symbol indicates that the summation only runs
over indices with $\vec l_1 \, a < \vec l_2 \, b$ and $\vec l_1^{\prime} \,
a^{\prime} < \vec l_2^{\prime} \, b^{\prime}$ to avoid double counting of
contributions.
The corresponding ansatz for~$\mat M^-(\vec k,
\omega)$ is formally identical to Eq.~(\ref{eq:ADC-form-expanded}) apart from the
changed occupation numbers of the
band indices~$i \to a$, $a \to i$, $b \to j$, $i^{\prime} \to a^{\prime}$,
$a^{\prime} \to i^{\prime}$, and $b^{\prime} \to j^{\prime}$. The summation
variables~$\vec l_1$, $\vec l_2$, $\vec l_1^{\prime}$, and $\vec l_2^{\prime}$
are crystal momentum vectors like~$\vec k$.
The quantities
$\mat M^{\pm}(\vec k,\omega)$, $\mat U^{\pm}(\vec k)$,
$\mat K^{\pm}(\vec k)$, and $\mat C^{\pm}(\vec k)$
are invariant under translations by an
arbitrary lattice vector, a property which is
taken into account for the internal summations in the
ADC~form~(\ref{eq:ADC-form-expanded}).

Explicit molecular ADC equations up to fourth order are given in
Ref.~\onlinecite{Schirmer:GF-83}. Here we show only the ADC equations
for the dynamic self-energy in terms of crystal (Bloch)
orbitals up to second order~[CO-ADC(2)]:
\begin{widetext}
  \begin{subeqnarray}
    \label{eq:k-CO-ADC2}
    \slabel{eq:k-CO-ADC2-U}
    U^+_{\vec k \, p \; ; \; \vec l_1 + \vec l_2  - \vec k \, i \;
      \vec l_1 \, a \; \vec l_2 \, b}
      &=& V^*_{\vec k \, p \;
      \vec l_1 + \vec l_2 - \vec k \, i \; [\vec l_1 \, a \; \vec l_2 \, b]} \>
      n_{\vec l_1 + \vec l_2 - \vec k \, i} \, \bar n_{\vec l_1 \, a}
      \, \bar n_{\vec l_2 \, b} \; , \\
    \slabel{eq:k-CO-ADC2-K}
    K^+_{\scriptstyle \vec l_1 + \vec l_2 - \vec k \, i \; \vec l_1 \, a \;
      \vec l_2 \, b \; ; \atop
      \scriptstyle \vec l_1^{\prime} + \vec l_2^{\prime} - \vec k \, i^{\prime} \;
      \vec l_1^{\prime} \, a^{\prime} \; \vec l_2^{\prime} \, b^{\prime}}
      &=& \delta_{\vec l_1 + \vec l_2 - \vec k \, i \; \vec l_1 \, a \;
      \vec l_2 \, a \; ; \atop
      \vec l_1^{\prime} + \vec l_2^{\prime} - \vec k \, i^{\prime} \;
      \vec l_1^{\prime} \, a^{\prime} \; \vec l_2^{\prime} \, b^{\prime}} \,
      (\varepsilon_{\vec l_1 \, a} + \varepsilon_{\vec l_2 \, b}
      - \varepsilon_{\vec l_1 + \vec l_2 - \vec k \, i}) \\
      && \qquad {} \times n_{\vec l_1 + \vec l_2 - \vec k \, i} \,
        \bar n_{\vec l_1 \, a} \, \bar n_{\vec l_2 \, b} \; , \nonumber \\
    \slabel{eq:k-CO-ADC2-C}
    C^+_{\scriptstyle \vec l_1 + \vec l_2 - \vec k \, i \; \vec l_1 \, a \;
      \vec l_2 \, b \; ; \atop
      \scriptstyle \vec l_1^{\prime} + \vec l_2^{\prime} - \vec k \, i^{\prime} \;
      \vec l_1^{\prime} \, a^{\prime} \; \vec l_2^{\prime} \, b^{\prime}}
      &=& 0  \; .
  \end{subeqnarray}
\end{widetext}
The equations for~$\mat U^-(\vec k)$, $\mat K^-(\vec k)$, and
$\mat C^-(\vec k)$ are formally identical to Eq.~(\ref{eq:k-CO-ADC2})
apart from the change of labels as for the adaptation of
Eq.~(\ref{eq:ADC-form-expanded}) to~$M_{pq}^-(\vec k, \omega)$
and the occupation numbers in Eqs.~(\ref{eq:k-CO-ADC2-U}) and
(\ref{eq:k-CO-ADC2-K}) which are in this
case~$\bar n_{\vec l_1 + \vec l_2 - \vec k \, a} \, n_{\vec l_1 \, i}
\, n_{\vec l_2 \, j}$.
Expression~(\ref{eq:k-CO-ADC2}) is derived in
Ref.~\onlinecite{Deleuze:SC-03} utilizing a Gaussian basis set
expansion of the Bloch orbitals.

The spectral representation~(\ref{eq:2h1p-spectral}) of the dynamic self-energy
is obtained from the ADC~form~(\ref{eq:ADC-form}) by solving the Hermitian
eigenvalue problem
\begin{equation}
  \label{eq:ADC-inversion}
  \begin{array}{rcl}
    & (\mat K^{\pm}(\vec k) + \mat C^{\pm}(\vec k)) \,
      \mat Y^{\pm}(\vec k) = \mat Y^{\pm}(\vec k) \,
      \mat \Omega^{\pm}(\vec k) & \\
    & \mat Y^{\pm \dagger}(\vec k) \, \mat Y^{\pm}(\vec k)
      = \unitmatrix \; . &
  \end{array}
\end{equation}
The vector notation in Eqs.~(\ref{eq:ADC-form}) and (\ref{eq:ADC-inversion})
is a convenient means to sum over all intermediate crystal momenta and band indices.
The diagonal matrix~$\mat \Omega^{\pm}(\vec k)$ contains the eigenvalues
of the secular matrix~$\mat K^{\pm}(\vec k) + \mat C^{\pm}(\vec k)$,
and $\mat Y^{\pm}(\vec k)$~denotes the eigenvector matrix.
The eigenvalues are the pole positions of the dynamic self-energy,
\ie, $\omega_n^{\pm}(\vec k) = \Omega^{\pm}_{nn}(\vec k)$, while
the Dyson amplitudes in Eq.~(\ref{eq:2h1p-spectral}) are obtained
via
\begin{equation}
  \label{eq:moddysontraf}
  m_p^{\pm,(n)}(\vec k) = \vec U_p^{\pm\dagger}(\vec k) \,
    \vec Y^{\pm,(n)}(\vec k) \; ,
\end{equation}
where the $n$-th column vector~$\vec Y^{\pm,(n)}(\vec k)$
of~$\mat Y^{\pm}(\vec k)$ is used here together with the adjoint
of~$\vec U_p^{\pm}(\vec k)$.
Note that,
unlike CO-ADC(2), in CO-ADC(3) (and higher orders)
the
eigenvectors~$\vec Y^{\pm,(n)}(\vec k)$ couple several
$2p1h$-, $2h1p$- or higher excited configurations
via the modified coupling amplitudes~(\ref{eq:k-CO-ADC2-C}) which are
nonzero in this case.

\subsection{Wannier representation}
\label{sec:WannierRepM}

We derive in the following an expansion of the dynamic self-energy
in terms of generalized Wannier orbitals; such orbitals are localized within
unit cells and allow one to apply cutoff
criteria inside a cell and between a cell and
other cells. This affords one the ability to apply fine-grained configuration
selection which is independent of the actual choice of the unit cell.
In particular, it enables the treatment of crystals
with large unit cells, not amenable to Bloch-orbital-based
approaches [Sec.~\ref{sec:onion}].~\cite{Fulde:EC-95,Fulde:WF-02,%
Ladik:QT-88,Ladik:PS-99}

The generalized Wannier transformation~\cite{Wannier:EE-37,%
Ashcroft:SSP-76,Ladik:QT-88,Callaway:QT-91,Marzari:ML-97,%
Ladik:PS-99,Zicovich:GM-01} and its inverse which mediate
between Bloch and Wannier orbitals are
\begin{subeqnarray}
  \label{eq:WannierTrafo}
  \slabel{eq:Wannier}
  w_{\vec R \, \varrho}(\vec r) &=& \frac{1}{\sqrt{N_0}} \,
    \Sum_{\vec k} \Sum_{p=1}^K {\cal U}_{p\varrho}(\vec k) \,
    \euler^{-\imag \vec k \vec R} \, \psi_{\vec k \, p}(\vec r) \; , \\
  \slabel{eq:InvWannier}
  \psi_{\vec k \, p}(\vec r) &=& \frac{1}{\sqrt{N_0}} \,
    \Sum_{\vec R} \Sum_{\varrho=1}^K {\cal U}^*_{p\varrho}(\vec k) \,
    \euler^{ \imag \vec k \vec R} \, w_{\vec R \, \varrho}(\vec r) \; ,
\end{subeqnarray}
where $K$~denotes the number of bands involved in the transformation
and $\mat{\cal U}(\vec k)$~is a suitable unitary matrix
as generated, \eg, by the localization procedures in
Refs.~\onlinecite{Marzari:ML-97,Zicovich:GM-01}.
Both Bloch and Wannier orbitals are normalized to
unity upon integration over the entire crystal which consists of $N_0$~unit cells.
Generally, we denote the Wannier orbital indices
that result from occupied bands with~$\kappa$, $\lambda$, $\mu$, $\nu$, \ldots$\,$,
while the Wannier orbital indices of unoccupied bands
are denoted by~$\alpha$, $\beta$, $\gamma$, $\delta$, \ldots$\,$.
Indices of Wannier orbitals from occupied or
unoccupied bands are referred to by~$\rho$, $\sigma$,
$\tau$, $\upsilon$, \ldots$\,$.

For an evaluation of the one-particle Green's function in terms
of the Feynman-Dyson perturbation series, it is essential that there is no
mixing between occupied and virtual Wannier orbitals;
otherwise Wick's theorem can no longer be
applied.~\cite{Mattuck:FD-76,Fetter:MP-71,Gross:MP-91} To
fulfill this requirement, the Wannier
transformation~(\ref{eq:Wannier}) is applied to the occupied and virtual
Bloch orbitals separately, yielding two independent unitary matrices,
one for the occupied bands and one for the virtual bands,
respectively.
Hence, $\mat {\cal U}(\vec k)$ in
Eq.~(\ref{eq:WannierTrafo}) is block-diagonal.

On the one hand, we want to evaluate the CO-ADC equations entirely in terms
of Wannier orbitals, in the so-called Wannier
or local representation;~\cite{Callaway:QT-91}
on the other hand, band structures are defined
with respect to the crystal momentum quantum number.
In what follows three different transformation schemes to switch between
the local representation and the crystal momentum
representation are applied
to the CO-ADC formalism and their physical and methodological
implications are discussed.

The dynamic self-energy~$M_{nm}^{\pm}(\vec k,\omega) \equiv
M_{\vec k \, n \; \vec k \, m}^{\pm}(\omega)$ depends on two
external Bloch orbitals
$\psi^{\dagger}_{\vec k \, n}(\vec r)$ and
$\psi_{\vec k \, m}(\vec r)$.
Hence, carrying out the inverse Wannier transformation~(\ref{eq:InvWannier})
for the external orbitals in Eq.~(\ref{eq:ADC-form-expanded})
leads to
\begin{equation}
  \label{eq:sigmawannier}
  \begin{array}{rcl}
    M_{pq}^{\pm}(\vec k,\omega) &=& {1 \over N_0} \Sum_{\varrho, \sigma}
      \Cal U_{p\varrho}(\vec k) \; \Cal U^*_{q\sigma}(\vec k) \\
    &&{} \times \Sum_{\vec R, \vec R^{\prime}}
      \euler^{\imag \vec k (\vec R^{\prime} - \vec R)} \;
      M_{\vec R \, \varrho \; \vec R^{\prime} \,
      \sigma}^{\pm}(\omega)
  \end{array}
\end{equation}
for the left-hand side of Eq.~(\ref{eq:ADC-form-expanded}) and a similar
expression for the right-hand side where the internal
summations~$\vec l_1$, $\vec l_2$, $\vec l_1^{\prime}$, and
$\vec l_2^{\prime}$ still run over the Brillouin zone.

In order to obtain a representation of the dynamic self-energy~$M_{nm}^{\pm}(\vec k,\omega)$
entirely in terms of Wannier orbitals, it is important to note
that, although matrix elements such as~$U^+_{\vec k \, p \; ; \; \vec l_1
+ \vec l_2  - \vec k \, i \; \vec l_1 \, a \; \vec l_2 \, b}$
in Eq.~(\ref{eq:ADC-form-expanded})
only depend on three independent crystal momenta, they
actually describe quantities which depend on four Bloch orbitals,
namely,~$\psi_{\vec k \, p}(\vec r)$,
$\psi_{\vec l_1 + \vec l_2  - \vec k \, i}(\vec r)$,
$\psi^{\dagger}_{\vec l_1 \, a}(\vec r)$, and
$\psi^{\dagger}_{\vec l_2 \, b}(\vec r)$
and thus a fourfold inverse
Wannier transformation~(\ref{eq:InvWannier}) has to be applied.
This transformation of Eq.~(\ref{eq:ADC-form-expanded})
to the Wannier representation yields
\begin{eqnarray}
  M_{\vec R \, \varrho \; \vec R^{\prime} \, \sigma}^+(\omega) &=& {\Sum_{
    \scriptstyle \vec g_1 \, \kappa, \; \vec g_2 \, \alpha, \; \vec g_3 \, \beta, \atop
    \scriptstyle \vec g_1^{\,\prime} \, \kappa^{\prime}\!\!, \; \vec g_2^{\,\prime}
    \, \alpha^{\prime}\!\!, \; \vec g_3^{\,\prime} \,
    \beta^{\prime}}}^{\hspace{-0.65cm}\prime\hspace{0.5cm}}
    {\check U}_{\vec R \, \varrho \; ; \; \vec g_1 \, \kappa \; \vec g_2 \,
    \alpha \; \vec g_3 \, \beta}^{+ \; *} \hspace{1.1cm} \nonumber \\
  \label{eq:Mwannier}
  &&{} \times
    (\omega \, \unitmatrix - \mat{\check K}^+ - \mat{\check C}^+)^{-1}_{
    \scriptstyle \vec g_1
    \, \kappa \; \vec g_2 \, \alpha \; \vec g_3 \, \beta \; ; \atop
    \scriptstyle \vec g_1^{\,\prime}
    \, \kappa^{\prime} \; \vec g_2^{\,\prime} \, \alpha^{\prime} \; \vec g_3^{\,\prime}
    \, \beta^{\prime}} \\
  &&{} \times {\check U}_{\vec R^{\prime} \, \sigma \; ; \; \vec g_1^{\,\prime} \, \kappa^{\prime} \;
    \vec g_2^{\,\prime} \, \alpha^{\prime} \; \vec g_3^{\,\prime} \,
    \beta^{\prime}}^+ \; , \nonumber
\end{eqnarray}
exploiting the unitarity of~$\mat{\Cal U}(\vec k)$ and the
basic relations~\cite{Ashcroft:SSP-76,Callaway:QT-91}
\begin{subeqnarray}
  \slabel{eq:orthonormality}
  \Sum_{\vec R} \euler^{\imag \vec k \vec R}
  &=& N_0 \, \delta_{\vec k, \, \vec 0} \; , \\
  \Sum_{\vec k} \euler^{\imag \vec k \vec R}
  &=& N_0 \, \delta_{\vec R, \, \vec 0} \; .
\end{subeqnarray}
The check accent on the quantities~$\mat{\check U}^+$,
$\mat{\check K}^+$, and $\mat{\check C}^+$
in Eq.~(\ref{eq:Mwannier}) indicates
that the sums run over lattice vectors and
Wannier orbital indices rather than crystal momenta and band
indices as in Eq.~(\ref{eq:ADC-form-expanded}).
The prime on the
summation symbol indicates that~$\vec g_2 \, \alpha < \vec g_3 \, \beta$
and $\vec g_2^{\,\prime} \, \alpha^{\prime} < \vec g_3^{\,\prime} \,
\beta^{\prime}$ must hold to avoid double counting of
contributions.
The corresponding equation for~$M_{\vec R \, \varrho
\; \vec R^{\prime} \, \sigma}^-(\omega)$ is formally identical to
Eq.~(\ref{eq:Mwannier}) apart from the changed occupation numbers~$\kappa
\to \alpha$, $\alpha \to \kappa$, $\beta \to \lambda$, $\kappa^{\prime} \to
\alpha^{\prime}$, $\alpha^{\prime} \to \kappa^{\prime}$, and $\beta^{\prime} \to
\lambda^{\prime}$.

One obviously Hermitian form for the dynamic self-energy emerging
from Eqs.~(\ref{eq:sigmawannier}) and (\ref{eq:Mwannier}) is given by
\begin{subeqnarray}
  \label{eq:Mcluster}
  \slabel{eq:Mclusterass}
  M_{pq}^{\pm}(\vec k,\omega) &=& \Sum_{\varrho, \sigma}
    \Cal U_{p\varrho}(\vec k) \; \Cal U^*_{q\sigma}(\vec k) \>
    \vec{\check U}_{\varrho}^{\pm}(\vec k)^{\dagger} \\
  \slabel{eq:Mclusteradd}
  &&{} \times (\omega \, \unitmatrix - \mat{\check K}^{\pm} - \mat{\check C}^{\pm})^{-1}
    \, \vec{\check U}_{\sigma}^{\pm}(\vec k) \; , \nonumber \\
  \vec{\check U}_{\sigma}^{\pm}(\vec k) &=& \frac{1}{\sqrt{N_0}} \sum_{\vec R} {\rm e}^{i
    \vec k \vec R} \; \vec{\check U}_{\vec R \, \sigma}^{\pm} \; .
\end{subeqnarray}
The vector notation in Eq.~(\ref{eq:Mcluster}) combines the six
internal summations over the intermediate lattice vectors
and Wannier orbital indices~$\vec g_1 \, \kappa$,
$\vec g_2 \, \alpha$, $\vec g_3 \, \beta$,
$\vec g_1^{\,\prime} \, \kappa^{\prime}$, $\vec g_2^{\,\prime} \, \alpha^{\prime}$,
and $\vec g_3^{\,\prime} \, \beta^{\prime}$
in Eq.~(\ref{eq:Mwannier}).
We refer to Eq.~(\ref{eq:Mcluster}) as the
\emph{supercell form} of CO-ADC.

The matrix~$\mat{\check K}^{\pm} + \mat{\check C}^{\pm}$ in
Eq.~(\ref{eq:Mclusterass}) does not explicitly dependend
on~$\vec k$ which implies that its
eigenvalues comprise all pole positions, \ie, pole positions for each $\vec k$-point.
Consequently, the resulting eigenvalues
also do not explicitly depend on~$\vec k$.
Multiplying with~$\vec{\check U}_{\varrho}^{\pm \, \dagger}(\vec k)$
from the left projects on the desired set of eigenvectors. Hence,
the translational symmetry among the $N_0$~unit cells of the crystal is
not exploited to reduce the size of the configuration
space which is unfavorable in conjunction
with the cutoff criteria discussed in Sec.~\ref{sec:onion}.
In practice, Eq.~(\ref{eq:Mcluster}) is applied to a molecular cluster
formed by $N_0$~unit cells of a crystal for which Born von K\'arm\'an
boundary conditions are enforced. As a consequence, only~$N_0$
discrete crystal momenta are in the Brillouin zone.
The above given transformation from the Wannier representation to
the crystal momentum representation
is simple and robust and can be used in conjunction with almost any
electron correlation method, but it suffers from finite-size effects
due to the, usually, small number of unit cells which are
considered in
practice; see, \eg, Ref.~\onlinecite{Foulkes:QMC-01}.

The translational symmetry of the dynamic
self-energy~(\ref{eq:Mwannier})
can be exploited by applying a translation
by~$-\vec R$ which removes the explicit dependence
of~$M_{\vec R \, \varrho \; \vec R^{\prime} \, \sigma}(\omega)$
on two external lattice
vectors such that it only depends on the difference of the lattice
vectors according to
\begin{equation}
  \label{eq:transdyn}
  \textstyle
  M_{\vec R \, \varrho \; \vec R^{\prime} \, \sigma}(\omega)
  = M_{\vec 0 \, \varrho \; \vec R^{\prime} - \vec R \, \sigma}(\omega)
  \equiv M_{\varrho\sigma} (\vec R^{\prime} - \vec R,\omega) \; .
\end{equation}
This facilitates to remove the lattice summation
over~$\vec R$ and the prefactor~$\frac{1}{N_0}$
in Eq.~(\ref{eq:sigmawannier}) which results in
(dropping the prime on~$\vec R^{\prime}$)
\begin{equation}
  \label{eq:sigmak}
  \begin{array}{rcl}
    M^{\pm}_{pq}(\vec k,\omega) &=& \Sum_{\varrho, \sigma} \Cal U_{p\varrho}(\vec k) \;
      \Cal U^*_{q\sigma}(\vec k) \\
    &&{} \times \Sum_{\vec R} {\rm e}^{i \vec k \vec R} \;
      M_{\varrho\sigma}(\vec R,\omega) \; .
  \end{array}
\end{equation}
To continue, we give up the idea of an ADC~form
for~$M_{\vec 0 \, \varrho \; \vec R \, \sigma}(\omega)$,
where the modified coupling amplitudes on the left and
on the right are related by Hermitian conjugation,
as is the case in Eq.~(\ref{eq:Mcluster}),
and arrive at
\begin{subeqnarray}
  \label{eq:M_k}
  M_{pq}^{\pm}(\vec k,\omega) &=& \Sum_{\varrho, \sigma}
    \Cal U_{p\varrho}(\vec k) \; \Cal U^*_{q\sigma}(\vec k) \>
    \vec{\tilde U}_{\vec 0 \, \varrho}^{\pm \dagger} \; , \\
  \slabel{eq:M_k_Dyson_Amp}
  &&{} \times (\omega \, \unitmatrix - \mat{\tilde K}^{\pm} -
    \mat{\tilde C}^{\pm})^{-1} \, \vec{\tilde U}_{\sigma}^{\pm}(\vec k)
    \nonumber \\
  \vec{\tilde U}_{\sigma}^{\pm}(\vec k) &=& \sum_{\vec R} \euler^{\imag
    \vec k \vec R} \; \vec{\tilde U}_{\vec R \, \sigma}^{\pm} \; ,
\end{subeqnarray}
where a slightly changed definition of the
transformed modified coupling
amplitude~$\vec{\tilde U}_{\sigma}^{\pm}(\vec k)
= \sqrt{N_0} \, \vec{\check U}_{\sigma}^{\pm}(\vec k)$ is employed
which is indicated by a tilde accent on the
quantities~$\mat{\tilde U}^{\pm} \equiv \mat{\check U}^{\pm}$,
$\mat{\tilde K}^{\pm} \equiv
\mat{\check K}^{\pm}$, and $\mat{\tilde C}^{\pm} \equiv \mat{\check C}^{\pm}$.
Because of its asymmetric nature, we denote Eq.~(\ref{eq:M_k})
as the \emph{semitransformed form}.

By exploiting the translational symmetry~(\ref{eq:transdyn}),
we remove the summation
over all translationally equivalent octuples consisting of
the two external Wannier orbitals~$w^{\dagger}_{\vec R
\, \varrho}(\vec r)$ and $w_{\vec R^{\prime} \, \sigma}(\vec r)$
of the dynamic self-energy~$M_{\vec R \, \varrho \;
\vec R^{\prime} \, \sigma}(\omega)$ and two $2p1h$- or
$2h1p$-configurations.
We arrive at a summation over~$\vec R^{\prime} -
\vec R$, $\varrho$, and $\sigma$
in conjunction with the remaining six internal lattice vectors and Wannier
orbital indices.
This formulation thus provides a far better starting point
for reducing the number of configurations by means of the
cutoff criteria of Sec.~\ref{sec:onion} than the supercell form.
Using the unsymmetric ADC~form~(\ref{eq:M_k}) implies that
the pole positions
still do not dependent explicitly on~$\vec k$.
Moreover, the modified coupling amplitudes~(\ref{eq:M_k_Dyson_Amp})
are, in contrast to Eq.~(\ref{eq:Mclusteradd}), no
longer related by Hermitian conjugation;
yet, for sure, the dynamic self-energy itself stays Hermitian.

This way of exploiting translational symmetry in conjunction with
the Wannier orbitals has already been applied
successfully before to devise \emph{ab initio} electron
correlation methods for crystals like the local Hamiltonian
approach~\cite{Grafenstein:VB-93,Grafenstein:VB-97,%
Albrecht:LO-98,Albrecht:AA-00,Bezugly:MC-04}
or the Green's function based method of Albrecht~\etal.~\cite{Albrecht:LO-01,%
Albrecht:LOB-02,Albrecht:LAI-02}

The internal structure of the dynamic self-energy
is not regarded in the two formulas~(\ref{eq:Mcluster})
and (\ref{eq:M_k}).
Expression~(\ref{eq:M_k}) results from Eqs.~(\ref{eq:sigmawannier})
and (\ref{eq:Mwannier}) by exploiting
the translational symmetry of~$\mat M^{\pm}(\vec k,\omega)$
in Eq.~(\ref{eq:transdyn}) and utilizing its exclusive dependence on
two external Wannier orbitals.
However, the translational symmetry of the
matrix~$\mat{\tilde K}^{\pm} + \mat{\tilde C}^{\pm}$
in Eq.~(\ref{eq:M_k}) can also be exploited additionally
which already was harnessed to derive Eq.~(\ref{eq:ADC-form-expanded})
from its original molecular orbital formulation.
We obtain
\begin{widetext}
  \begin{subeqnarray}
    \label{eq:M_k_hermitian}
    \slabel{eq:m_k_h_M}
    M_{pq}^+(\vec k,\omega) &=&  \Sum_{\varrho, \sigma}
      \Cal U_{p\varrho}(\vec k) \; \Cal U^*_{q\sigma}(\vec k) \> \Sum_{
      \scriptstyle \vec g_1, \vec g_2, \atop \scriptstyle \vec g_1^{\,\prime}, \vec g_2^{\,\prime}}
      \Sum_{\scriptstyle \kappa, \alpha, \beta, \atop
            \scriptstyle \kappa^{\prime}\!, \alpha^{\prime}\!,
      \beta^{\prime}}
      \bar U_{\varrho \; ; \; \vec g_1 \; \vec g_2 \; \kappa \, \alpha \, \beta}^{+ \; *}
      (\vec k) \nonumber \\
    &&{} \times (\omega \, \unitmatrix - \mat{\bar K}^+(\vec k)
      - \mat{\bar C}^+(\vec k))^{-1}_{\scriptstyle \vec g_1 \; \vec g_2 \; \kappa \, \alpha \, \beta \; ;
      \atop \scriptstyle \vec g_1^{\,\prime} \; \vec g_2^{\,\prime} \; \, \kappa^{\prime} \,
      \alpha^{\prime} \, \beta^{\prime}} \\
    &&{} \times \bar U_{\sigma \; ; \; \vec g_1^{\,\prime} \; \vec g_2^{\,\prime} \;
      \kappa^{\prime} \, \alpha^{\prime} \, \beta^{\prime}}^+(\vec k) \; ,
      \nonumber \\
    \slabel{eq:m_k_h_U}
    \bar U_{\sigma \; ; \; \vec g_1^{\,\prime} \; \vec g_2^{\,\prime} \; \kappa^{\prime} \,
      \alpha^{\prime} \, \beta^{\prime}}^+(\vec k) &=&
      \sum_{\vec R} \euler^{\imag \vec k \vec R} \; \bar U^+_{\vec R \, \sigma \; ; \;
      \vec 0 \, \kappa^{\prime} \; \vec g_1^{\,\prime} \, \alpha^{\prime} \;
      \vec g_2^{\,\prime} \, \beta^{\prime}} \; , \\
    \slabel{eq:m_k_h_KC}
    (\mat{\bar K}^+(\vec k) + \mat{\bar C}^+(\vec k))_{\scriptstyle \vec g_1 \; \vec g_2 \; \kappa
      \, \alpha \, \beta \; ; \atop \scriptstyle \vec g_1^{\,\prime} \; \vec g_2^{\,\prime} \; \,
      \kappa^{\prime} \, \alpha^{\prime} \, \beta^{\prime}} &=&
      \sum_{\vec R} \euler^{\imag \vec k \vec R} \;
      (\mat{\bar K}^+ + \mat{\bar C}^+)_{\scriptstyle \vec 0 \, \kappa \; \vec g_1 \, \alpha \; \vec g_2 \,
      \beta \; ; \atop \scriptstyle
      \vec R \, \kappa^{\prime} \; \vec g_1^{\,\prime} + \vec R \,
      \alpha^{\prime} \; \vec g_2^{\,\prime} + \vec R \, \beta^{\prime}} \; ,
  \end{subeqnarray}
\end{widetext}
where the identities~$\mat{\bar U}^+ \equiv \mat{\check U}^+$,
$\mat{\bar K}^+ \equiv \mat{\check K}^+$, and
$\mat{\bar C}^+ \equiv \mat{\check C}^+$ hold.
The corresponding relations for~$\mat M^-(\vec k,\omega)$ are formally
identical to Eq.~(\ref{eq:M_k_hermitian}) apart from the changed occupation
numbers~$\kappa \to \alpha$,
$\alpha \to \kappa$, $\beta \to \lambda$, $\kappa^{\prime} \to \alpha^{\prime}$,
$\alpha^{\prime} \to \kappa^{\prime}$, and $\beta^{\prime} \to \lambda^{\prime}$.
Alternatively, we can derive Eq.~(\ref{eq:M_k_hermitian}) by inserting
the inverse Wannier transformation~(\ref{eq:InvWannier}) into the CO-ADC
equations for Bloch orbitals~(\ref{eq:ADC-form-expanded})
which already are fully adapted to
translational symmetry.
We refer to Eq.~(\ref{eq:M_k_hermitian}) as the \emph{fully
translational symmetry adapted form}.

The modified coupling amplitudes~(\ref{eq:m_k_h_U}) are constructed
by considering only Wannier orbitals relative to an origin
cell (or reference cell) where the hole
with index~$\kappa$ is assumed to reside. The
external orbital index~$\sigma$ of a modified coupling
amplitude can be viewed to represent
a Bloch orbital~$\psi_{\vec k \, \sigma}(\vec r)$. This
orbital interacts with the
$2p1h$-configurations
that are pinned with one lattice
vector to the origin cell and extend with the two
remaining lattice vectors~$\vec g_1$ and $\vec g_2$
over up to two different unit cells. Alternatively,
one can use the translational symmetry
of the~$\bar U^+_{\vec R \, \sigma \; ; \; \vec 0 \,
\kappa^{\prime} \; \vec g_1^{\,\prime} \, \alpha^{\prime} \;
\vec g_2^{\,\prime} \, \beta^{\prime}}$
to arrive at
\begin{equation}
  \label{eq:m_k_h_U_trans}
  \bar U_{\sigma \; ; \; \vec g_1^{\,\prime} \; \vec g_2^{\,\prime} \;
    \kappa^{\prime} \, \alpha^{\prime} \, \beta^{\prime}}^+(\vec k)
  = \Sum_{\vec R} \euler^{-\imag \vec k \vec R} \; \bar U^+_{\vec 0 \,
    \sigma \; ; \; \vec R \, \kappa^{\prime} \; \vec g_1^{\,\prime}
    + \vec R \, \alpha^{\prime} \; \vec g_2^{\,\prime}
    + \vec R \, \beta^{\prime}} \; .
\end{equation}
Now the external orbital is~$w_{\vec 0 \, \sigma}(\vec r)$ which is independent
of~$\vec R$ and consequently Eq.~(\ref{eq:m_k_h_U_trans}) is
interpreted to describe the interaction of
a Wannier orbital in the origin cell with $2p1h$-configurations centered
around all lattice vectors~$\vec R$ which are combined to give an intermediate
$2p1h$-configuration with a total crystal momentum~$-\vec k$.

The matrix~$\mat{\bar K}^+ + \mat{\bar C}^+$ in
Eq.~(\ref{eq:m_k_h_KC}) describes the coupling among the
$2p1h$-configurations. One of the lattice vectors
of the first triple of Wannier orbitals
is pinned to the origin cell, and the two other lattice
vectors~$\vec g_1$ and $\vec g_2$ are offsets
to it. The remaining three lattice vectors belong to a Wannier
orbital in an arbitrary cell~$\vec R$ and two further
Wannier orbitals in the cells~$\vec g_1^{\,\prime} + \vec R$ and $\vec g_2^{\,\prime}
+ \vec R$ relative to the former. Obviously the entire Eq.~(\ref{eq:m_k_h_KC})
can be interpreted to describe the interaction of a
$2p1h$-configuration of Wannier orbitals centered around the origin cell with
another $2p1h$-configuration
with crystal momentum~$\vec k$.
Due to the full exploitation of translational symmetry, also
of the intermediate $2p1h$-configurations, the
fully translational symmetry adapted form yields the
smallest configuration space in conjunction with the cutoff
criteria of Sec.~\ref{sec:onion}.

In order to give explicit CO-ADC expressions for the matrices showing up
in Eq.~(\ref{eq:M_k_hermitian}),
we transform the Hamiltonian~(\ref{eq:Bloch_Hamilton})
with the help of formula~(\ref{eq:InvWannier}) to
the Wannier representation.
A similar decomposition as in
Eq.~(\ref{eq:Bloch_Hamilton}) is chosen for a subsequent
application of perturbation theory:
\begin{equation}
  \label{eq:Wannier_Hamilton}
  \renewcommand{\arraystretch}{1.5}%
  \begin{array}{rcl}
    \hat H &=& \hat H^{\rm WF}_0 + \hat H^{\rm WF}_1 \; , \\
    \hat H^{\rm WF}_0 &=& \Sum_{\vec R \, \varrho} \varepsilon_{\vec R \,
      \varrho} \, \hat c_{\vec R \, \varrho}^{\dagger} \,
      \hat c_{\vec R \, \varrho} \; , \\
    \hat H^{\rm WF}_1 &=& \Sum_{\vec R_1 \, \varrho, \  \vec R_2 \, \sigma}
      W_{\vec R_1 \, \varrho \; \vec R_2 \, \sigma} \> \hat c_{\vec R_1
      \, \varrho}^{\dagger} \, \hat c_{\vec R_2 \, \sigma} \\
    &&{} + \frac{1}{2} \Sum_{\vec R_1 \, \varrho, \ \vec R_2 \, \sigma,
      \atop  \vec R_3 \, \tau, \  \vec R_4 \, \upsilon} V_{\vec R_1 \,
      \varrho \; \vec R_2 \, \sigma \; \vec R_3 \, \tau \; \vec
      R_4 \, \upsilon} \\
    &&{} \times \hat c_{\vec R_1 \, \varrho}^{\dagger} \, \hat c_{\vec R_2
      \, \sigma}^{\dagger} \, \hat c_{\vec R_4 \, \upsilon}
      \, \hat c_{\vec R_3 \, \tau} \; ,
  \end{array}
  \renewcommand{\arraystretch}{1}%
\end{equation}
with~$\varepsilon_{\vec R \, \varrho} \equiv F_{\vec R \, \varrho \; \vec R \,
\varrho} = F_{\vec 0 \, \varrho \; \vec 0 \, \varrho}$
being the diagonal elements of the Fock matrix~$\mat F$. Here
$W_{\vec R_1 \, \varrho \; \vec R_2 \, \sigma} = -\Sum_{\vec R \, \tau}
V_{\vec R_1 \, \varrho \; \vec R \, \tau \; [\vec R_2 \, \sigma \; \vec R
\, \tau]} \, n_{\vec R \, \tau} + (1 - \delta_{\vec R_1 \, \varrho \;
\vec R_2 \, \sigma}) \, F_{\vec R_1 \, \varrho \; \vec R_2 \, \sigma}$
comprises the negative of the Hartree-Fock potential and
also the off-diagonal elements of the Fock matrix.
Note that~$\mat F$
is block-diagonal
and thus the Wannier orbitals are still separated into occupied and
virtual Wannier orbitals.

Using the inverse generalized
Wannier transformation~(\ref{eq:InvWannier})---as
is done throughout this article---corresponds to the use of the
full Fock matrix as zeroth-order Hamiltonian. In addition, a
perturbative expansion of the off-diagonal elements in the Fock matrix is
applied in succession of the inverse Wannier transformation. This conduct
is equivalent to employing the partitioning of the Fock matrix
introduced in Eq.~(\ref{eq:Wannier_Hamilton}).

The one-electron interaction is mediated by the
off-diagonal Fock matrix elements, and the two-electron
interaction is represented by the two-electron integrals;
let the one-electron interaction
be treated using Feynman diagrams up to $m$-th~order
and account for diagrams which exclusively involve two-electron
interactions of highest order~$n$. This separation
facilitates to take higher-order terms into account
for the slowlier convergent series; the resulting
scheme is denoted by CO-ADC($m$,$n$).

The CO-ADC(2,2) approximation for Wannier orbitals
is obtained similarly to the case of the CO-ADC(2)
approximation for Bloch orbitals~(\ref{eq:k-CO-ADC2})
and reads
\begin{widetext}
  \begin{subeqnarray}
    \label{eq:R-CO-ADC2}
    \slabel{eq:m_R_h_U}
    \bar U^+_{\vec R \, \varrho \; ; \; \vec g_1 \, \kappa \; \vec g_2 \, \alpha \;
      \vec g_3 \, \beta} &=& V^*_{\vec R \, \varrho \;
      \vec g_1 \, \kappa \; [\vec g_2 \, \alpha \; \vec g_3 \, \beta]} \>
      n_{\vec g_1 \, \kappa} \, \bar n_{\vec g_2 \, \alpha} \,
      \bar n_{\vec g_3 \, \beta} \; , \\
    \slabel{eq:m_R_h_K}
    \bar K^+_{\vec g_1 \, \kappa \; \vec g_2 \, \alpha \; \vec g_3 \, \beta \; ; \atop
      \vec g_1^{\,\prime} \, \kappa^{\prime} \; \vec g_2^{\,\prime} \, \alpha^{\prime} \;
      \vec g_3^{\,\prime} \, \beta^{\prime}}
      &=& \delta_{\vec g_1 \, \kappa \; \vec g_2 \, \alpha \; \vec g_3 \, \beta \; ; \atop
      \vec g_1^{\,\prime} \, \kappa^{\prime} \; \vec g_2^{\,\prime} \, \alpha^{\prime} \;
      \vec g_3^{\,\prime} \, \beta^{\prime}} \,
      (\varepsilon_{\vec g_2 \, \alpha} + \varepsilon_{\vec g_3 \, p_3}
      - \varepsilon_{\vec g_1 \, \kappa}) \>
      n_{\vec g_1 \, \kappa} \, \bar n_{\vec g_2 \, \alpha} \,
      \bar n_{\vec g_3 \, \beta} \; , \\
    \slabel{eq:m_R_h_C}
    \bar C^+_{\vec g_1 \, \kappa \; \vec g_2 \, \alpha \; \vec g_3 \, \beta \; ; \atop
      \vec g_1^{\,\prime} \, \kappa^{\prime} \; \vec g_2^{\,\prime} \, \alpha^{\prime} \;
      \vec g_3^{\,\prime} \, \beta^{\prime}} &=& 0 \; .
  \end{subeqnarray}
\end{widetext}
The equations for~$\mat{\bar U}^-$, $\mat{\bar K}^-$, and $\mat{\bar C}^-$ are
formally identical to Eq.~(\ref{eq:R-CO-ADC2}) upon modifying
the labels according to the adaptation of Eq.~(\ref{eq:M_k_hermitian})
to~$M_{pq}^-(\vec k,\omega)$ and replacing the
occupation numbers in Eqs.~(\ref{eq:m_R_h_U}) and
(\ref{eq:m_R_h_K}) by~$\bar n_{\vec g_1 \, \alpha}
\, n_{\vec g_2 \, \kappa} \, n_{\vec g_3 \, \lambda}$.
Up to second order, the perturbative treatment of the off-diagonal elements
of the Fock matrix
does not introduce any
new diagrams for the dynamic self-energy and thus $\mat{\bar K}^{\pm}
+ \mat{\bar C}^{\pm}$ remains diagonal.~\cite{Cederbaum:TA-77,Santra:CAP-02}
However, in a formulation which is invariant under unitary
transformations in the occupied and virtual space,
this is no longer the case because the full Fock matrix
needs to be chosen as the zeroth-order
Hamiltonian in place of~$\hat H^{\rm WF}_0$
then.~\cite{Buth:MC-05,Birkenheuer:ND-04}

\section{Static self-energy}
\label{sec:static}
\subsection{Crystal momentum representation}
\label{sec:crystalS}

The static self-energy has not been determined so far. It is represented
by all those diagrams in the diagrammatic expansion of the self-energy
where the external points of the diagrams correspond to equal
times.~\cite{Cederbaum:TA-77} In a strict second-order treatment of
electron correlations, using Bloch orbitals~(\ref{eq:Bloch_Hamilton}),
the static self-energy is zero because the first static self-energy
diagrams arise the earliest in third order.
Moreover, it turns out that the diagrammatic series for the static self-energy
does not converge reasonably in many cases.~\cite{Schirmer:SE-89} However, a
self-consistent solution is possible~\cite{Schirmer:GF-83,Niessen:CM-84}
utilizing~\cite{Ethofer:SP-69,Winter:CE-72,Cederbaum:TA-77}
\begin{equation}
  \label{eq:staticS}
  \begin{array}{rcl}
    \Sigma_{pq}^{\infty}(\vec k) &=& W_{pq}(\vec k) + \Sum_{\vec k^{\prime}}
      \Sum_{r,s} V_{\vec k \, p \; \vec k^{\prime} \, r \; [\vec k \, q \;
      \vec k^{\prime} s]} \\
    &&{} \times \biggl [ \frac{1}{2\pi\imag} \oint G_{sr}(\vec k^{\prime},
      \omega) \differential \omega \biggr ] \; ,
  \end{array}
\end{equation}
here given in terms of Bloch orbitals. The contour integration
in Eq.~(\ref{eq:staticS}) runs along the real axis and closes in the
upper complex $\omega$-plane, hence enclosing only the poles of the
advanced Green's function~$\mat G^-(\vec k^{\prime}, \omega)$
in Eq.~(\ref{eq:p-decomp}).

The self-consistent solution of Eq.~(\ref{eq:staticS}) is
computationally expensive. Yet a stable and efficient Dyson expansion
method~(DEM) for determining the static self-energy is devised in
Ref.~\onlinecite{Schirmer:SE-89}. There, the first two terms of the Dyson
expansion~(\ref{eq:DysonExp}) are inserted into Eq.~(\ref{eq:staticS}). The
term~$W_{pq}(\vec k)$ in Eq.~(\ref{eq:staticS}) cancels the result of the contour
integration over the free Green's function~$G_{sr}^0(\vec k^{\prime}, \omega)$.
Carrying out the contour integration over the product of the two free Green's
functions and the static self-energy
yields
\begin{widetext}
  \begin{subeqnarray}
    \label{eq:statick}
    \slabel{eq:kstaticimplicit}
    \Sigma_{pq}^{\infty}(\vec k) &=& \Sum_{\vec k^{\prime}} \Sum_{r,
      s} A_{\vec k \, p \; \vec k \, q, \; \vec k^{\prime} \, s \; \vec k^{\prime} \, r}
      \, \Sigma_{sr}^{\infty}(\vec k^{\prime}) + b_{pq}(\vec k) \; , \\
    \slabel{eq:kA}
    A_{\vec k \, p \; \vec k \, q, \; \vec k^{\prime} \, s \; \vec k^{\prime} \, r}
      &=& V_{\vec k \, p \; \vec k^{\prime} \, r \; [\vec k \, q \; \vec k^{\prime} \, s]}
      \> (  \varepsilon_{\vec k^{\prime} \, r}
          - \varepsilon_{\vec k^{\prime} \, s})^{-1} \;
      \biggl [       n_{\vec k^{\prime} \, r} \, \bar n_{\vec k^{\prime} \, s}
              - \bar n_{\vec k^{\prime} \, r} \,      n_{\vec k^{\prime} \, s}
      \biggr ] \; , \\
    \slabel{eq:kbnm}
    b_{pq}(\vec k) &=& \Sum_{\vec k^{\prime}} \Sum_{r,
      s} V_{\vec k \, p \; \vec k^{\prime} \, r \; [\vec k \, q \;
      \vec k^{\prime} \, s]} \, Q_{rs}(\vec k^{\prime}) \; , \\
    \slabel{eq:kQqp}
    Q_{rs}(\vec k^{\prime}) &=& \frac{1}{2\pi\imag} \oint G^0_{ss}(\vec k^{\prime},\omega)
      \, M_{sr}(\vec k^{\prime}, \omega) \,
      G^0_{rr}(\vec k^{\prime},\omega) \differential \omega \; .
  \end{subeqnarray}
The~$Q_{rs}(\vec k^{\prime})$ can now be determined by
inserting the spectral representation of the dynamic
self-energy~(\ref{eq:2h1p-spectral})
into Eq.~(\ref{eq:kQqp}).
Carrying out the contour integration yields
  \begin{subeqnarray}
    \label{eq:Qkl_Mlk-k}
    Q_{rs}(\vec k^{\prime}) &=& Q^+_{rs}(\vec k^{\prime})
                                + Q^-_{rs}(\vec k^{\prime}) \; , \\
    Q^+_{rs}(\vec k^{\prime}) &=& \sum_{n \in \{N+1\}} m_s^{+,(n)}(\vec k^{\prime})
      \> m_r^{+,(n)*}(\vec k^{\prime}) \, \biggl[
             \frac{    -n_{\vec k^{\prime} \, r} \,      n_{\vec k^{\prime} \, s}}
                  {(\varepsilon_{\vec k^{\prime} \, r} - \omega_n^+(\vec k^{\prime}))
                   (\varepsilon_{\vec k^{\prime} \, s} - \omega_n^+(\vec k^{\prime}))}
      \nonumber \\
      &&{} + \frac{     n_{\vec k^{\prime} \, r} \, \bar n_{\vec k^{\prime} \, s}}
             {(\varepsilon_{\vec k^{\prime} \, r} - \varepsilon_{\vec k^{\prime} \, s})
              (\varepsilon_{\vec k^{\prime} \, r} - \omega_n^+(\vec k^{\prime}))}
           - \frac{\bar n_{\vec k^{\prime} \, r} \,      n_{\vec k^{\prime} \, s}}
             {(\varepsilon_{\vec k^{\prime} \, r} - \varepsilon_{\vec k^{\prime} \, s})
              (\varepsilon_{\vec k^{\prime} \, s} - \omega_n^+(\vec k^{\prime}))}
      \biggr] \; , \\
    Q^-_{rs}(\vec k^{\prime}) &=& \sum_{n \in \{N-1\}} m_s^{-,(n)}(\vec k^{\prime})
      \> m_r^{-,(n)*}(\vec k^{\prime}) \, \biggl[
             \frac{\bar n_{\vec k^{\prime} \, r} \, \bar n_{\vec k^{\prime} \, s}}
             {(\varepsilon_{\vec k^{\prime} \, r} - \omega_n^-(\vec k^{\prime}))
              (\varepsilon_{\vec k^{\prime} \, s} - \omega_n^-(\vec k^{\prime}))}
      \nonumber \\
      &&{} - \frac{\bar n_{\vec k^{\prime} \, r} \,      n_{\vec k^{\prime} \, s}}
             {(\varepsilon_{\vec k^{\prime} \, r} - \varepsilon_{\vec k^{\prime} \, s})
              (\varepsilon_{\vec k^{\prime} \, r} - \omega_n^-(\vec k^{\prime}))}
           + \frac{     n_{\vec k^{\prime} \, r} \, \bar n_{\vec k^{\prime} \, s}}
             {(\varepsilon_{\vec k^{\prime} \, r} - \varepsilon_{\vec k^{\prime} \, s})
              (\varepsilon_{\vec k^{\prime} \, s} - \omega_n^-(\vec k^{\prime}))}
      \biggr] \; .
  \end{subeqnarray}
\end{widetext}

In order to evaluate Eq.~(\ref{eq:Qkl_Mlk-k}),
from now on, we resort to
the ADC~form of the dynamic
self-energy in terms of Bloch orbitals~(\ref{eq:ADC-form-expanded}).
By affixing a bar accent to the
quantities~$\mat U^{\pm}(\vec k)$,
$\mat K^{\pm}(\vec k)$, and $\mat C^{\pm}(\vec k)$ in
all following equations of this subsection, they become formally identical
to the ones obtained with the fully translational
symmetry-adapted form~(\ref{eq:M_k_hermitian}).

The direct usage of Eq.~(\ref{eq:Qkl_Mlk-k}) is in practice very
time consuming because the~$\mat K^{\pm}(\vec k^{\prime})
+ \mat C^{\pm}(\vec k^{\prime})$ matrix has to be diagonalized fully~(\ref{eq:ADC-inversion})
to obtain the pole
positions~$\omega_n^{\pm}(\vec k^{\prime})$
of the spectral representation~(\ref{eq:2h1p-spectral}).
For the CO-ADC(2)
approximation of the
dynamic self-energy in crystal momentum representation~(\ref{eq:k-CO-ADC2}),
however, $\mat K^{\pm}(\vec k^{\prime}) +
\mat C^{\pm}(\vec k^{\prime})$~remains
diagonal. This leads
to~$m_r^{\pm,(n)*}(\vec k^{\prime}) = U_{\vec k^{\prime} \, r, n}^{\pm}$
and $\omega_n^{\pm}(\vec k^{\prime}) = K_{nn}^{\pm}(\vec k^{\prime})$ where the
$n$-th~eigenvector of~$\mat K^{\pm}(\vec k^{\prime}) + \mat C^{\pm}(\vec k^{\prime})$
is unity for a single $2p1h$- or $2h1p$-configuration, respectively, and zero otherwise. In general,
the~$Q_{rs}(\vec k^{\prime})$ are determined efficiently from
Eq.~(\ref{eq:Qkl_Mlk-k}) by the inversion method or by a single-vector Lanczos
diagonalization to circumvent a full diagonalization of~$\mat K^{\pm}(\vec k^{\prime})
+ \mat C^{\pm}(\vec k^{\prime})$.~\cite{Schirmer:SE-89}

The inversion method has been found to be more
efficient~\cite{Schirmer:SE-89} than the single-vector Lanczos
diagonalization, so we concentrate
on the former. We define auxiliary vectors
$\vec V_r^{\pm}(\vec k^{\prime})$ as the solution
of the inhomogeneous systems of linear equations
\begin{widetext}
\begin{equation}
  \label{eq:Vr}
  (\varepsilon_{\vec k^{\prime} \, r} \unitmatrix - \mat K^{\pm}(\vec k^{\prime})
  - \mat C^{\pm}(\vec k^{\prime})) \, \vec V_r^{\pm}(\vec k^{\prime})
  = \vec U_r^{\pm}(\vec k^{\prime}) \; .
\end{equation}
Inserting the eigenvector matrix~$\mat Y^{\pm}(\vec k^{\prime})$
from Eq.~(\ref{eq:ADC-inversion})
into Eq.~(\ref{eq:Vr}), one obtains, by using
formula~(\ref{eq:moddysontraf}),
\begin{equation}
  \vec V_r^{\pm}(\vec k^{\prime}) = \mat Y^{\pm}(\vec k^{\prime}) \,
  (\varepsilon_{\vec k^{\prime} \, r} \unitmatrix - \mat \Omega^{\pm}
  (\vec k^{\prime}))^{-1} \, \vec m_r^{\pm\dagger}(\vec k^{\prime}) \; ,
\end{equation}
which reveals the usefulness of the new vectors and allows us
to rewrite Eq.~(\ref{eq:Qkl_Mlk-k}) as follows:
\begin{subeqnarray}
  \label{eq:Qrs_Vr}
  Q_{rs}^+(\vec k^{\prime}) &=& -\vec V_s^{+\dagger}(\vec k^{\prime}) \,
    \vec V_r^+(\vec k^{\prime}) \, n_{\vec k^{\prime} \, r} \, n_{\vec
    k^{\prime} \, s} + (\varepsilon_{\vec k^{\prime} \, r}
    - \varepsilon_{\vec k^{\prime} \, s})^{-1} \nonumber \\
  &&{} \times [ \vec U_s^{+\dagger}(\vec k^{\prime}) \,
    \vec V_r^+(\vec k^{\prime}) \, n_{\vec k^{\prime} \, r} \,
    \bar n_{\vec k^{\prime} \, s} - \vec V_s^{+\dagger}(\vec k^{\prime})
    \, \vec U_r^+(\vec k^{\prime}) \, \bar n_{\vec k^{\prime} \, r} \,
    n_{\vec k^{\prime} \, s} ] \; , \\
  Q_{rs}^-(\vec k^{\prime}) &=& \vec V_s^{-\dagger}(\vec k^{\prime}) \,
    \vec V_r^-(\vec k^{\prime}) \, \bar n_{\vec k^{\prime} \, r} \, \bar
    n_{\vec k^{\prime} \, s} - (\varepsilon_{\vec k^{\prime} \, r}
    - \varepsilon_{\vec k^{\prime} \, s})^{-1} \nonumber \\
  &&{} \times [ \vec U_s^{-\dagger}(\vec k^{\prime}) \,
                \vec V_r^ -        (\vec k^{\prime}) \,
    \bar n_{\vec k^{\prime} \, r} \, n_{\vec k^{\prime} \, s}
    - \vec V_s^{-\dagger}(\vec k^{\prime})
    \, \vec U_r^-(\vec k^{\prime}) \, n_{\vec k^{\prime} \, r} \,
    \bar n_{\vec k^{\prime} \, s} ] \; .
\end{subeqnarray}
\end{widetext}

We have reduced the problem of determining the~$Q_{rs}^{\pm}(\vec k^{\prime})$
to the problem of determining the~$\vec V_r^{\pm}(\vec k^{\prime})$.
Inspecting
Eqs.~(\ref{eq:Vr}) and (\ref{eq:Qrs_Vr}),
we note that we have to solve for the~$\vec V_r^+(\vec k^{\prime})$
in the large $2p1h$-configuration space but only for the usually small
set of all occupied orbitals.
Conversely, the~$\vec V_r^-(\vec k^{\prime})$
have to be calculated in the small $2h1p$-configuration space but
for the usually large set of all virtual orbitals.

The diagonal parts~$\varepsilon_{\vec k^{\prime} \, r} \unitmatrix -
\mat K^{\pm}(\vec k^{\prime})$ in Eq.~(\ref{eq:Vr})
are, at least, of a magnitude of twice the Hartree-Fock band gap,
implying usually a diagonal dominance of the full matrices in
Eq.~(\ref{eq:Vr}). Therefore, a solution by Jacobi
iterations~\cite{Schirmer:SE-89,Golub:MC-96} is suggested:
\begin{equation}
  \begin{array}{rcl}
    \vec V_r^{\pm, (0)}(\vec k^{\prime}) &=& (\varepsilon_{\vec k^{\prime}
    \, r} \unitmatrix - \mat K^{\pm}(\vec k^{\prime}))^{-1} \,
    \vec U_r^{\pm}(\vec k^{\prime}) \; , \\
    \vec V_r^{\pm, (n)}(\vec k^{\prime}) &=& \vec V_r^{\pm, (0)}(\vec k^{\prime})
      + (\varepsilon_{\vec k^{\prime} \, r} \unitmatrix - \mat
      K^{\pm}(\vec k^{\prime}))^{-1} \\
    &&{} \qquad \times \mat C^{\pm}(\vec k^{\prime}) \, \vec V_r^{\pm,
      (n-1)}(\vec k^{\prime}) \; ,
  \end{array}
\end{equation}
which turns out to converge rapidly.

The inhomogeneous linear system of equations~(\ref{eq:kstaticimplicit})
can now be solved for~$\Sigma_{pq}^{\infty}(\vec k)$ by a matrix
inversion~\cite{Niessen:CM-84}:
\begin{equation}
  \label{eq:getsigmak}
  \vec \Sigma^{\infty} = (\unitmatrix - \mat A)^{-1} \, \vec b \; ,
\end{equation}
where~$\Sigma_i^{\infty}$, $b_i$, and $A_{ij}$ are composed by numerating
the compound indices~$(p, q, \vec k)$ and $(s, r, \vec k^{\prime})$
in~$\Sigma_{pq}^{\infty}(\vec k)$, $b_{pq}^{\infty}(\vec k)$, and
$A_{\vec k \, p \; \vec k \, q, \; \vec k^{\prime} \, s \; \vec k^{\prime}
\, r}$ by integer numbers~$i$ and $j$, respectively. Inspecting
Eq.~(\ref{eq:kA}), we note that $\mat A$ couples only to the components of
$\Sigma_{sr}^{\infty}(\vec k^{\prime})$ where the $s,r$~indices denote
a combination of particle-hole- or hole-particle-orbitals. As all quantities in Eq.~(\ref{eq:getsigmak})
are Hermitian, we can further restrict the band indices to~$p \leq q$.
As soon as the number of $\vec k$-points becomes large, the solution of
Eq.~(\ref{eq:getsigmak}) gets cumbersome as the dimension of the system of
linear equations is given by the number of entries of the upper triangle (including
the diagonal entries) of the static self-energy matrix times the number
of $\vec k$-points. Thus an iterative linear equations solver has to be
employed.~\cite{Golub:MC-96}

\subsection{Wannier representation}
\label{sec:WannierRepS}

The static self-energy can also be determined directly in Wannier
representation.
Having expressed the static self-energy in terms of Wannier orbitals,
we obtain the static self-energy in crystal momentum representation by
\begin{equation}
  \label{eq:SigmakR}
  \Sigma_{pq}^{\infty}(\vec k) = \Sum_{\varrho, \sigma} \Cal
  U_{p\varrho}(\vec k) \; \Cal U^*_{q\sigma}(\vec k) \Sum_{\vec R}
  \euler^{\imag \vec k \vec R} \, \Sigma_{\varrho\sigma}^{\infty}
  (\vec R) \; .
\end{equation}
To find an approximation for~$\Sigma_{\varrho\sigma}^{\infty}(\vec R)$,
we insert the inverse
Wannier transformation~(\ref{eq:InvWannier}) and its Hermitian conjugate
into Eq.~(\ref{eq:statick}) and arrive at
\begin{subeqnarray}
  \label{eq:GetSigma}
  \slabel{eq:staticinhomo}
  \Sigma_{\varrho\sigma}^{\infty}(\vec R) &=& \Sum_{\vec R^{\prime}}
    \Sum_{\tau,\upsilon} \Bigl[ \Sum_{\vec g} A_{\vec 0 \, \varrho \;
    \vec R \, \sigma, \; \vec g \, \upsilon \; \vec g
    + \vec R^{\prime} \, \tau} \Bigr] \,
    \Sigma_{\upsilon\tau}^{\infty}(\vec R^{\prime}) \\
  &&{} + b_{\varrho\sigma}(\vec R) \; , \hspace{5.3cm} \nonumber \\
  \slabel{eq:bnm_Qqp}
  b_{\varrho\sigma}(\vec R) &=& \Sum_{\vec R^{\prime}} \Sum_{\tau,\upsilon}
    \Bigl[ \Sum_{\vec g} V_{\vec 0 \, \varrho \; \vec g +
    \vec R^{\prime} \, \tau \; [\vec R \,
    \sigma \; \vec g \, \upsilon]} \Bigr] \, Q_{\tau\upsilon}
    (\vec R^{\prime}) \; .
\end{subeqnarray}
Note that the translational symmetry of~$A_{\vec 0 \, \varrho \;
\vec R \, \sigma, \; \vec g \, \upsilon \; \vec g
+ \vec R^{\prime} \, \tau}$
could only be utilized
once to remove one lattice summation such that two lattice summations
(one over $\vec R^{\prime}$ and one over $\vec g$) show up in
Eq.~(\ref{eq:GetSigma}) while only one summation over~$\vec k^{\prime}$
is necessary in Eq.~(\ref{eq:statick}). There is no first-order
contribution to the static self-energy in Eq.~(\ref{eq:GetSigma}) because
the static self-energy used here
evolves from an
inverse Wannier transformation of the self-energy obtained in
M\o{}ller-Plesset partition~\cite{Szabo:MQC-89,McWeeny:MQM-92,Helgaker:MES-00}
and a subsequent
perturbative expansion with respect to the off-diagonal elements of
the Fock matrix.

For the remaining equations~(\ref{eq:kA}) and
(\ref{eq:kQqp}), we
assume the partition~(\ref{eq:Wannier_Hamilton}) of
the Hamiltonian in Wannier representation, to obtain
\begin{subeqnarray}
  \label{eq:staticcoeff}
  A_{\vec 0 \, \varrho \; \vec R \, \sigma, \; \vec g
    \, \upsilon \; \vec g + \vec R^{\prime} \, \tau}
  &=& V_{\vec 0 \, \varrho \; \vec g + \vec R^{\prime}
    \, \tau \; [\vec R \, \sigma \; \vec g \, \upsilon]} \nonumber \\
  &&{} \times (\varepsilon_{\vec g + \vec R^{\prime} \, \tau}
    - \varepsilon_{\vec g \, \upsilon})^{-1} \\
  &&{} \times (
    n_{\vec g + \vec R^{\prime} \, \tau} \, \bar n_{\vec g \, \upsilon}
    - \bar n_{\vec g + \vec R^{\prime} \, \tau} \, n_{\vec g \, \upsilon}
    ) \; , \qquad \nonumber \\
  \slabel{eq:Qtauupsilon}
  Q_{\tau\upsilon}(\vec R^{\prime}) &\equiv&
  Q_{\vec R^{\prime} \, \tau \; \vec 0 \, \upsilon} =
    \frac{1}{2\pi\imag} \oint
    G_{\upsilon\upsilon}^0(\vec 0,\omega) \\
  &&{} \times M_{\upsilon\tau}(\vec R^{\prime},\omega) \,
    G_{\tau\tau}^0(\vec 0,\omega) \differential \omega \; , \nonumber
\end{subeqnarray}
with the free Green's function in Wannier representation
given by
\begin{equation}
  \begin{array}{rcl}
    G^0_{\tau\upsilon}(\vec R^{\prime} - \vec R, \omega) &\equiv&
    G^0_{\vec R \, \tau \; \vec R^{\prime} \, \upsilon}(\omega) \\
    &=&
      \delta_{\vec R \, \tau, \; \vec R^{\prime} \, \upsilon} \>
      \Bigl[ \frac{\bar n_{\vec R \, \tau}}
                  {\omega - \varepsilon_{\vec R \, \tau} + \imag \eta}
           + \frac{n_{\vec R \, \tau}}
                  {\omega - \varepsilon_{\vec R \, \tau} - \imag \eta}
      \Bigr] \; .
  \end{array}
\end{equation}

To determine the~$Q_{\tau\upsilon}(\vec R^{\prime})$, we introduce the
spectral representation of the dynamic self-energy in Wannier representation
[the analog of Eq.~(\ref{eq:2h1p-spectral})]
\begin{eqnarray}
  M_{\upsilon\tau}(\vec R^{\prime},\omega) &=& \sum_{n \in \{N+1\}}
    \frac{\check m_{\upsilon}^{+,(n)}(\vec 0) \>
          \check m_{\tau}^{+,(n)*}(\vec R^{\prime})}
         {\omega - \check\omega^+_n + \imag \eta} \nonumber \\
  \label{eq:2h1p-spectral-R}
  &&{}+ \sum_{n \in \{N-1\}}
    \frac{\check m_{\upsilon}^{-,(n)}(\vec 0) \>
          \check m_{\tau}^{-,(n)*}(\vec R^{\prime})}
         {\omega - \check\omega^-_n - \imag \eta} \\
  &=& M_{\upsilon\tau}^+(\vec R^{\prime},\omega)
    + M_{\upsilon\tau}^-(\vec R^{\prime},\omega)
    \hspace{1.5cm} \; , \nonumber
\end{eqnarray}
which is obtained from the ADC~form of the dynamic self-energy in Wannier
representation~(\ref{eq:Mwannier}) by letting~$\vec R \to \vec 0$
and
diagonalizing~$\mat{\check K}^{\pm} + \mat{\check C}^{\pm}$
analogously to Eq.~(\ref{eq:ADC-inversion}).
Since translational
symmetry cannot be exploited in
Eq.~(\ref{eq:Mwannier})
as~$\vec R^{\prime}$---in contrast to~$\vec k$---is not a good quantum
number, this leads to the redundancies that have
already been mentioned in conjunction with the supercell form of the
dynamic self-energy in Eq.~(\ref{eq:Mcluster});
namely, the full configuration space of $2p1h$- and
$2h1p$-configurations from the $N_0$~unit cells of the
crystal is necessary here.
However, the cutoff criterion of Sec.~\ref{sec:onion},
nevertheless, can be applied.

The~$Q_{\tau\upsilon}(\vec R^{\prime})$ can now be
determined by inserting
Eq.~(\ref{eq:2h1p-spectral-R}) into Eq.~(\ref{eq:Qtauupsilon}).
Carrying out the contour integration yields
\begin{widetext}
\begin{subeqnarray}
  \label{eq:Qkl_Mlk}
  Q_{\tau\upsilon}(\vec R^{\prime}) &=& Q_{\tau\upsilon}^+(\vec R^{\prime})
    + Q_{\tau\upsilon}^-(\vec R^{\prime}) \; , \\
  Q^+_{\tau\upsilon}(\vec R^{\prime}) &=& \sum_{n \in \{N+1\}}
    \check m_{\upsilon}^{+,(n)}(\vec 0)
    \> \check m_{\tau}^{+,(n)*}(\vec R^{\prime}) \, \biggl[
           \frac{    -n_{\vec R^{\prime} \, \tau} \,      n_{\vec 0 \, \upsilon}}
           {(\varepsilon_{\vec R^{\prime} \, \tau} - \check \omega_n^+)
            (\varepsilon_{\vec 0 \, \upsilon} - \check \omega_n^+)} \nonumber \\
    &&{} + \frac{     n_{\vec R^{\prime} \, \tau} \, \bar n_{\vec 0 \, \upsilon}}
           {(\varepsilon_{\vec R^{\prime} \, \tau} - \varepsilon_{\vec 0 \, \upsilon})
            (\varepsilon_{\vec R^{\prime} \, \tau} - \check \omega_n^+)}
         - \frac{\bar n_{\vec R^{\prime} \, \tau} \,      n_{\vec 0 \, \upsilon}}
           {(\varepsilon_{\vec R^{\prime} \, \tau} - \varepsilon_{\vec 0 \, \upsilon})
            (\varepsilon_{\vec 0 \, \upsilon} - \check \omega_n^+)} \biggr] \; , \\
  Q^-_{\tau\upsilon}(\vec R^{\prime}) &=& \sum_{n \in \{N-1\}}
    \check m_{\upsilon}^{-,(n)}(\vec 0)
    \> \check m_{\tau}^{-,(n)*}(\vec R^{\prime}) \, \biggl[
           \frac{\bar n_{\vec R^{\prime} \, \tau} \, \bar n_{\vec 0 \, \upsilon}}
           {(\varepsilon_{\vec R^{\prime} \, \tau} - \check \omega_n^-)
            (\varepsilon_{\vec 0 \, \upsilon} - \check \omega_n^-)} \nonumber \\
    &&{} - \frac{\bar n_{\vec R^{\prime} \, \tau} \,      n_{\vec 0 \, \upsilon}}
           {(\varepsilon_{\vec R^{\prime} \, \tau} - \varepsilon_{\vec 0 \, \upsilon})
            (\varepsilon_{\vec R^{\prime} \, \tau} - \check \omega_n^-)}
         + \frac{     n_{\vec R^{\prime} \, \tau} \, \bar n_{\vec 0 \, \upsilon}}
           {(\varepsilon_{\vec R^{\prime} \, \tau} - \varepsilon_{\vec 0 \, \upsilon})
            (\varepsilon_{\vec 0 \, \upsilon} - \check \omega_n^-)} \biggr] \; .
\end{subeqnarray}
We proceed as before by defining the auxiliary vectors
\begin{equation}
  (\varepsilon_{\vec R^{\prime} \, \varrho} \unitmatrix
    - \mat{\check K}^{\pm} - \mat{\check C}^{\pm}) \,
    \vec V_{\varrho}^{\pm}(\vec R^{\prime}) =
    \vec U_{\varrho}^{\pm}(\vec R^{\prime}) \; ,
\end{equation}
setting~$\vec U_{\varrho}^{\pm}(\vec R)
\equiv \vec{\check U}_{\vec R \, \varrho}^{\pm}$, and
rewrite Eq.~(\ref{eq:Qkl_Mlk}):
\begin{subeqnarray}
  \label{eq:Qkl-R}
  Q_{\tau\upsilon}^+(\vec R^{\prime}) &=& - \vec V_{\upsilon}^{+\dagger}(\vec 0)
    \, \vec V_{\tau}^+(\vec R^{\prime})
    \, n_{\vec R^{\prime} \, \tau} \, n_{\vec 0 \, \upsilon}
    + (\varepsilon_{\vec R^{\prime} \, \tau} - \varepsilon_{\vec 0 \, \upsilon})^{-1}
    \nonumber \\
    &&{} \times [ \vec U_{\upsilon}^{+\dagger}(\vec 0) \,
    \vec V_{\tau}^+(\vec R^{\prime}) \, n_{\vec R^{\prime} \, \tau} \,
    \bar n_{\vec 0 \, \upsilon} - \vec V_{\upsilon}^{+\dagger}(\vec 0)
    \, \vec U_{\tau}^+(\vec R^{\prime}) \, \bar n_{\vec R^{\prime} \, \tau} \,
    n_{\vec 0 \, \upsilon} ] \; , \\
  Q_{\tau\upsilon}^-(\vec R^{\prime}) &=& \vec V_{\upsilon}^{-\dagger}(\vec 0)
    \, \vec V_{\tau}^-(\vec R^{\prime})
    \, \bar n_{\vec R^{\prime} \, \tau} \, \bar n_{\vec 0 \, \upsilon}
    - (\varepsilon_{\vec R^{\prime} \, \tau} - \varepsilon_{\vec 0 \, \upsilon})^{-1}
    \nonumber \\
    &&{} \times [ \vec U_{\upsilon}^{-\dagger}(\vec 0) \,
    \vec V_{\tau}^-(\vec R^{\prime}) \, \bar n_{\vec R^{\prime} \, \tau} \,
    n_{\vec 0 \, \upsilon} - \vec V_{\upsilon}^{-\dagger}(\vec 0)
    \, \vec U_{\tau}^-(\vec R^{\prime}) \, n_{\vec R^{\prime} \, \tau} \,
    \bar n_{\vec 0 \, \upsilon} ] \; .
\end{subeqnarray}
\end{widetext}
The algorithm for Jacobi iterations now reads
\begin{equation}
  \begin{array}{rcl}
    \vec V_{\varrho}^{\pm, (0)}(\vec R^{\prime}) &=& (\varepsilon_{\vec
      R^{\prime} \, \varrho} \unitmatrix - \mat{\check K}^{\pm})^{-1}
      \, \vec U_{\varrho}^{\pm}(\vec R^{\prime}) \; , \\
    \vec V_{\varrho}^{\pm, (n)}(\vec R^{\prime}) &=& \vec V_{\varrho}^{\pm,
      (0)}(\vec R^{\prime}) + (\varepsilon_{\vec R^{\prime} \, \varrho}
      \unitmatrix - \mat{\check K}^{\pm})^{-1} \\
    &&\qquad \times \mat{\check C}^{\pm} \, \vec V_{\varrho}^{\pm, (n-1)}
      (\vec R^{\prime}) \; ,
  \end{array}
\end{equation}
and the inhomogeneous linear system of equations~(\ref{eq:staticinhomo}) can
be solved for~$\Sigma_{\varrho\sigma}^{\infty}(\vec R)$ by a matrix
inversion~\cite{Niessen:CM-84}:
\begin{equation}
  \label{eq:getsigmaR}
  \vec{\check\Sigma}^{\infty} = (\unitmatrix -
  \mat{\check A})^{-1} \, \vec{\check b} \; ,
\end{equation}
where a check accent is affixed to indicate that Wannier
orbitals rather than Bloch orbitals are used.
We employ a mapping of the compound indices~$(\varrho, \sigma, \vec R)$
and $(\upsilon, \tau, \vec R^{\prime})$ to the
integer numbers~$i$ and $j$, respectively,
for the quantities~$\check\Sigma_i^{\infty}$,
$\check b_i$, and $\check A_{ij}$.
Note that $\mat \Sigma^{\infty}(\vec R)$ is not Hermitian.
Hence, the full static self-energy matrix has to be included
in~$\vec{\check\Sigma}^{\infty}$ explicitly for several lattice
vectors~$\vec R$. However, the property
$\mat \Sigma^{\infty \> \dagger}(\vec R) = \mat \Sigma^{\infty}(-\vec R)$
holds which can be utilized to reduce the number
of equations in formula~(\ref{eq:getsigmaR}).
Nevertheless, in many cases, the linear system of equations needs
to be solved iteratively.~\cite{Golub:MC-96}

\section{Dyson equation}
\label{sec:poledyson}

Having determined approximations for the static and the
dynamic self-energy in terms of Wannier
orbitals in Eqs.~(\ref{eq:SigmakR}) and
(\ref{eq:M_k_hermitian}), respectively,
we can finally determine the positions and strengths
of the poles of the one-particle Green's function from
Eq.~(\ref{eq:solveDysonFormal}). [The pole positions and strengths
of the Green's function, using Bloch orbitals to represent the
self-energy~(\ref{eq:statick}) and (\ref{eq:ADC-form-expanded}), can be
obtained following a nearly identical line of argument.]
To this end, the small Hermitian eigenvalue problem
\begin{equation}
  \label{eq:selfconsistent}
  (\mat{\varepsilon}(\vec k) + \mat \Sigma^{\infty}(\vec k)
    + \mat M(\vec k,\omega) ) \, \vec{\check x}\I{G}(\vec k, \omega)
    = \omega \, \vec{\check x}\I{G}(\vec k, \omega) \; ,
\end{equation}
with $\mat \varepsilon(\vec k)$~being the diagonal matrix of Bloch orbital
energies and $\vec{\check x}\I{G}(\vec k, \omega)$~denoting eigenvectors,
has to be solved self-consistently, \ie, such that the
energy~$\omega$ entering the dynamic self-energy~$\mat M(\vec k,
\omega)$ is identical to the resulting eigenvalue~$\omega$.

To avoid the unitary matrices~$\mat{\cal U}(\vec k)$ and
$\mat{\cal U}^{\dagger}(\vec k)$ in
Eqs.~(\ref{eq:m_k_h_M}) and (\ref{eq:SigmakR}) arising
in the following expressions,
we multiply Eq.~(\ref{eq:selfconsistent}) from the left
with~$\mat{\cal U}^{\dagger}(\vec k)$ and
insert~$\mat{\cal U}(\vec k) \,
\mat{\cal U}^{\dagger}(\vec k) = \unitmatrix$
between the right parenthesis and the eigenvector~$\vec{\check x}\I{G}$
on the left-hand side of the formula.
Defining the translational symmetry-adapted
Fock matrix~$\bar F_{\varrho\sigma}(\vec k) = \Sum_{\vec R} \euler^{\imag
\vec k \vec R} \, F_{\varrho\sigma}(\vec R)$,
one arrives at
\begin{equation}
  \label{eq:fixedomega}
  (\mat{\bar F}(\vec k) + \mat{\bar \Sigma}^{\infty}(\vec k) + \mat{\bar M}(\vec k,\omega))
  \, \vec x\I{G}(\vec k, \omega) = \omega \, \vec x\I{G}(\vec k, \omega) \; ,
\end{equation}
where the new quantities~$\bar F_{\varrho\sigma}(\vec k)$, $\bar \Sigma_{\varrho\sigma}^{\infty}(\vec k)$,
$\bar M_{\varrho\sigma}(\vec k,\omega)$, and $x_{{\rm G}, \varrho}(\vec k, \omega)$
are related to the old ones~$\varepsilon_{pq}(\vec k)$, $\Sigma_{pq}^{\infty}(\vec k)$,
$M_{pq}(\vec k,\omega)$, and $\check x_{{\rm G}, p}(\vec k, \omega)$
by~$\mat\varepsilon(\vec k) = \mat{\Cal U}(\vec k) \,
\mat{\bar F}(\vec k) \, \mat{\Cal U}^{\dagger}(\vec k)$,
$\mat\Sigma^{\infty}(\vec k) = \mat{\Cal U}(\vec k) \,
\mat{\bar\Sigma}^{\infty}(\vec k) \,
\mat{\Cal U}^{\dagger}(\vec k)$,
$\mat M(\vec k,\omega) = \mat{\Cal U}(\vec k) \,
\mat{\bar M}(\vec k,\omega) \,
\mat{\Cal U}^{\dagger}(\vec k)$,
and
$\vec{\check x}\I{G}(\vec k, \omega)
= \mat{\Cal U}(\vec k) \,
\vec x\I{G}(\vec k, \omega)$.

Inserting the ADC~form of the dynamic
self-energy~(\ref{eq:m_k_h_M}) into
Eq.~(\ref{eq:fixedomega}) and defining
\begin{equation}
  \label{eq:fixeddynamic}
  \vec x^{\pm}(\vec k, \omega) = (\omega \, \unitmatrix
  - \mat{\bar K}^{\pm}(\vec k) - \mat{\bar C}^{\pm}(\vec k))^{-1}
  \, \mat{\bar U}^{\pm}(\vec k) \, \vec x\I{G}(\vec k, \omega) 
\end{equation}
yields the
following form of the eigenvalue problem~(\ref{eq:fixedomega}):
\begin{equation}
  \label{eq:unfixomega}
  \begin{array}{rcl}
    (\mat{\bar F}(\vec k) &\!+\!& \mat{\bar \Sigma}^{\infty}(\vec k)) \,
      \vec x\I{G}(\vec k, \omega) + \mat{\bar U}^{+\dagger}(\vec k) \,
      \vec x^+(\vec k, \omega) \\
    &&{}\quad + \mat{\bar U}^{-\dagger}(\vec k) \, \vec x^-(\vec k, \omega)
      = \omega \, \vec x\I{G}(\vec k, \omega) \; . \nonumber
  \end{array}
\end{equation}
Rewriting Eq.~(\ref{eq:fixeddynamic}) as
\begin{equation}
  \label{eq:evsub}
  \begin{array}{rcl}
    \mat{\bar U}^{\pm}(\vec k) \, \vec x\I{G}(\vec k, \omega)
      &\!+\!& (\mat{\bar K}^{\pm}(\vec k) + \mat{\bar C}^{\pm}(\vec k))
      \, \vec x^{\pm}(\vec k, \omega) \\
    &&{} = \omega \, \vec x^{\pm}(\vec k, \omega) \; ,
  \end{array}
\end{equation}
it becomes evident that Eqs.~(\ref{eq:unfixomega}) and (\ref{eq:evsub})
can be recast as a joint Hermitian eigenvalue problem
\begin{widetext}
\renewcommand{\arraystretch}{1.5}%
\begin{eqnarray}
  \label{eq:solvepADC}
  & \mat B(\vec k) \mat X(\vec k) = \mat X(\vec k) \mat E(\vec k),
  \qquad \mat X^{\dagger}(\vec k) \, \mat X(\vec k) = \unitmatrix \; ,
  & \nonumber \\[1.5ex]
  & \mat B(\vec k) = \left(
  \begin{array}{ccc}
    \mat{\bar F}(\vec k) + \mat{\bar \Sigma}^{\infty}(\vec k)
    & \mat{\bar U}^{+\dagger}(\vec k) & \mat{\bar U}^{-\dagger}(\vec k) \\
    \mat{\bar U}^+(\vec k) & \mat{\bar K}^+(\vec k) + \mat{\bar C}^+(\vec k) & \mat 0 \\
    \mat{\bar U}^-(\vec k) & \mat 0 & \mat{\bar K}^-(\vec k) + \mat{\bar C}^-(\vec k)
  \end{array}
  \right) \; . &
\end{eqnarray}
\renewcommand{\arraystretch}{1}%
\end{widetext}
The new matrix~$\mat B(\vec k)$ is called the \emph{band
structure matrix}.
It has to be diagonalized for several $\vec k$-points
yielding eigenvalues~$e_n(\vec k) = (\mat E(\vec k))_{nn}$
and eigenvectors~$\vec X_n(\vec k) =
(\vec x\I{G}(\vec k, e_n(\vec k))\transpose, \;
\vec x^+(\vec k, e_n(\vec k))\transpose, \; \vec x^-(\vec k,
e_n(\vec k))\transpose)\transpose$.
Formula~(\ref{eq:solvepADC}) is similar to the result for
molecules.~\cite{Schirmer:GF-83}
One can envisage that the diagonalization of the
subblocks~$\mat{\bar K}^{\pm}(\vec k) + \mat{\bar C}^{\pm}(\vec k)$
in~$\mat B(\vec k)$ is carried out first
by solving Eq.~(\ref{eq:ADC-inversion});
afterwards, the full eigenvalue problem~(\ref{eq:solvepADC}) is
treated with all subblocks modified
accordingly.~\cite{Schirmer:GF-83,Weikert:BL-96}
Therefore, via Eq.~(\ref{eq:ADCsuminfinity}),
infinitely many proper self-energy diagrams are summed
before the result is put in the Dyson
equation~(\ref{eq:Dyson}) to sum all improper self-energy diagrams
that derive from those contained in the ADC~form.

The spectral representation of the one-particle Green's
function~(\ref{eq:p-spectrum}) reads, in terms
of the eigenpairs of the band structure
matrix~(\ref{eq:solvepADC}),
\begin{equation}
  \label{eq:spectral_CO-ADC}
  G_{pq}(\vec k, \omega) = (\omega \, \unitmatrix - \mat B(\vec k))^{-1}_{pq}
    = \Sum_n \frac{\bar x_p^{(n)}(\vec k) \>
    \bar x_q^{(n)*}(\vec k)}{\omega - e_n(\vec k)} \; ,
\end{equation}
where $\bar x_p^{(n)}(\vec k) = (\mat X(\vec k))_{pn}$~denotes the
transition amplitude of the $n$-th state. It is given by
either the~$x_p^{(n)}(\vec k)$ or the~$y_p^{(n)}(\vec k)$ in Eq.~(\ref{eq:transamp})
whether it belongs to an $N - 1$ or an $N + 1$~particle state.

In some cases, the external Wannier orbitals of the
dynamic self-energy are restricted to a single cell
of the crystal, \ie, $\mat M(\vec k, \omega)
\approx \mat M(\vec R = \vec 0, \omega)$, the so-called
one-lattice-site approximation for the dynamic self-energy.
Note, however, that the $2p1h$- and $2h1p$-configurations
in neighboring unit cells are not
entirely excluded. The $\vec k$-dependence, in this
approximation, is solely due to the
one-particle matrix~$\mat{\bar F}(\vec k) +
\mat{\bar\Sigma}^{\infty}(\vec k)$.
The static self-energy can also be evaluated in one-lattice-site
approximation. Then the $\vec k$-dependence is exclusively due
to~$\mat{\bar F}(\vec k)$ and this approximation to the CO-ADC
equations becomes similar to the dynamical mean-field
theory.~\cite{Georges:DM-96}

\section{Configuration Selection}
\label{sec:onion}

All equations derived so far contain certain infinite lattice sums
that run over the entire crystal forming the configuration space. To
obtain a meaningful approximation of the one-particle Green's function,
the convergence of those lattice sums must be granted. Sun and
Bartlett~\cite{Sun:CL-97} study the convergence properties of the
lattice sums in several quantum chemical correlation methods for
ground states and exited states of crystals, based on the
M\o{}ller-Plesset partition of the Hamiltonian~(\ref{eq:Bloch_Hamilton}).
Following the reasoning of Ref.~\onlinecite{Sun:CL-97}, the lattice
sums in CO-ADC in crystal momentum representation converge.
Moreover, the equations in local representation
also converge as they evolve from the former
by means of
an inverse Wannier transformation~(\ref{eq:InvWannier}).

Although the convergence of lattice sums is granted principally, we
have to devise an algorithm for their proper
truncation;~\cite{Fulde:EC-95} \ie, a dynamical building of the
configuration space is required which means to meet
a chosen accuracy in the lattice sums for a given crystal.
The appropriate truncation of lattice sums is an
essential ingredient of all \emph{ab initio} methods for crystals
and corresponds to a configuration selection procedure which ensures that the
configuration space is sufficiently large for the desired
accuracy of the band structure but still sufficiently small to be tractable
on present-day computers. In molecular physics, configuration selection
has been introduced in the context of the
configuration interaction
method~\cite{Buenker:CS-74}
which suffers from an exponentially growing configuration space with respect to
the number of atoms in molecules. The techniques discussed in
Ref.~\onlinecite{Buenker:CS-74} are not applicable here, as they
are designed with the configuration interaction method in mind,
but there are analogs for crystals which are similar in spirit.

The incremental scheme for ground states of
crystals~\cite{Stoll:CS-92,Stoll:CD-92,Stoll:CG-92,Fulde:EC-95,%
Fulde:WF-02,Willnauer:QC-04,Buth:BS-04,Buth:MC-05}
is based on Wannier orbitals. The total configuration
space of a crystal is partitioned into certain subsets which are used in a
subsequent calculation of the correlation energy to yield the so-called
energy increments.
Frequently, explicit manual configuration selection is
applied.
This allows the partition of the configuration space to be chosen
following chemical intuition.
Yet the situation is much more cumbersome for band structures because
the number of distinct excited states grows with the number of
correlated electrons, \ie, the number of Bloch orbitals and
Wannier orbitals which are considered in the sums in
Eqs.~(\ref{eq:ADC-form-expanded})
and (\ref{eq:M_k_hermitian}), respectively.
In the local Hamiltonian
approach, the number of states to be calculated is restricted
to a fixed number which is treated in an incremental
way.~\cite{Grafenstein:VB-93,Grafenstein:VB-97,Albrecht:LO-98,
Albrecht:AA-00,Bezugly:MC-04} Albrecht~\etal{}~\cite{Albrecht:LO-01,
Albrecht:LAI-02} ensure the convergence of an incremental series
for the self-energy for~$\omega = 0$, a criterion which is independent
of the number of states actually described.

We devise here a configuration selection procedure that is
perfectly adapted to the structure of the CO-ADC scheme.
For the equations in terms of Bloch orbitals~(\ref{eq:statick})
and (\ref{eq:ADC-form-expanded}),
$\vec k$-points are chosen harnessing Born von K\'arm\'an
boundary conditions which lead to a net
of equidistant $\vec k$-points.~\cite{Ashcroft:SSP-76,Callaway:QT-91}
In this case, configuration selection means
choosing a sufficiently dense net in conjunction with
a cutoff criterion which is similar to the one for CO-ADC in terms
of Wannier orbitals~(\ref{eq:SigmakR}) and (\ref{eq:M_k_hermitian})
discussed in the next paragraph.
Yet beforehand, one should interpret
the impact of the number of $\vec k$-points used for
Brillouin zone integration.
Using a given net of $N_0$ $\vec k$-points
to carry out a Wannier transformation~(\ref{eq:Wannier}),
CO-ADC in terms of Bloch orbitals can
be immediately represented and analyzed in terms of the supercell
form~(\ref{eq:Mcluster}).
However, one should emphasize that the redundancies of this
representation---which is utilized here for interpretation
only---are not present in the Bloch orbital formulation of CO-ADC.
The supercell form affords a tight-binding interpretation:
within the supercell, the number of distinct neighbors to the
origin cell is restricted by the volume of the supercell.
Distinct interaction terms among the cells, \ie, the Fock matrix
and two-electron integrals, are thus also
resticted and
the number of $\vec k$-points
employed can thus be understood to imply the number of
nearest-neighbor cells treated in the interaction terms.
These arguments generalize the tight-binding
arguments given, \eg, in Ref.~\onlinecite{Monkhorst:SP-76},
in conjunction with a single Brillouin zone integration
of a crystal-momentum-dependent function to the
multi-dimensional case.

Configuration selection in crystal momentum representation
is not intuitive and does not allow a fine-grained
selection of configurations within unit cells and between
an origin cell and its neighbor cells. For crystals with
large unit cells this is a significant restriction.
To obtain a suitable cutoff criterion for CO-ADC in terms
of Wannier orbitals~(\ref{eq:SigmakR}) and
(\ref{eq:M_k_hermitian}), we evaluate the
second-order diagram of the retarded dynamic
self-energy~\cite{Cederbaum:TA-77,Schirmer:GF-83,Buth:MC-05}
and examine the summand therein:
\begin{equation}
  \label{eq:selfsecond}
  {V_{\vec 0 \, \varrho \; \vec g_1 \, \kappa \; [ \vec g_2 \, \alpha \; \vec
  g_3 \, \beta]} \, V_{\vec R \, \sigma \; \vec g_1 \, \kappa \; [ \vec g_2 \, \alpha
  \; \vec g_3 \, \beta]}^* \over
  \omega - \varepsilon_{\vec g_2 \, \alpha} - \varepsilon_{\vec g_3 \, \beta}
  + \varepsilon_{\vec g_1 \, \kappa}} \; n_{\vec g_1 \, \kappa} \;
  \bar n_{\vec g_2 \, \alpha} \, \bar n_{\vec g_3 \, \beta} \; .
\end{equation}
The most delocalized occupied orbitals are the valence orbitals.
Therefore, the occupied valence orbitals couple to the most important
$2p1h$-configurations. Hence, $\omega$ can be assumed to be of the order
of the band gap. The denominator in Eq.~(\ref{eq:selfsecond})
is regarded to be constant to a reasonable degree.
It is thus
considered as being part of the cutoff threshold. Therefore,
if~$V_{\vec 0 \, \varrho \;
\vec g_1 \, \kappa \; [ \vec g_2 \, \alpha \; \vec g_3 \, \beta]} \,
V_{\vec R \, \sigma \; \vec g_1 \, \kappa \; [ \vec g_2 \, \alpha
\; \vec g_3 \, \beta]}^* \; n_{\vec g_1 \, \kappa} \;
\bar n_{\vec g_2 \, \alpha} \, \bar n_{\vec g_3 \, \beta}$ is above a certain cutoff
threshold for some combination of~$\vec 0 \, \varrho$ and $\vec R \, \sigma$,
we include the corresponding $2p1h$-test-configuration~$(\vec g_1
\, \kappa, \vec g_2 \, \alpha, \vec g_3 \, \beta)$ in the configuration
space. If the product is small for all external Wannier orbitals of the
self-energy, $(\vec g_1 \, \kappa, \vec g_2 \, \alpha,
\vec g_3 \, \beta)$ is neglected completely by setting
all two-electron integrals
containing this configuration exactly to zero. By this, only a
finite range of the residual Coulomb interaction, which is not
treated in Hartree-Fock approximation, is explored.
The same arguments hold for the selection of $2h1p$-configurations for
the evaluation of the advanced dynamic self-energy,
if Eq.~(\ref{eq:selfsecond}) is adapted in the same
way as Eq.~(\ref{eq:R-CO-ADC2}) is adapted to~$\mat{\bar U}^-$,
$\mat{\bar K}^-$, and $\mat{\bar C}^-$.

Convergence of the configuration space, \ie, the number of unit cells
contributing configurations to be considered in the cutoff
criterion~(\ref{eq:selfsecond}), is expected to be sufficiently quick.
Van der Waals dispersion interactions which are present both
in the ground state of the $N$~particle system and
the states of the $N \pm 1$~particle system
are
nearly equal
and thus almost
cancel in band structure calculations.
However, the effect of the extra Coulomb charge that
arises in $N \pm 1$~particle states
in contrast to the $N$~particle ground state
is long range.
The potential of an extra point charge falls off
like~$V(r) = \pm 1 / (\varepsilon \, r)$,
where $\varepsilon$~is the dielectric constant of the solid.
This causes an overall polarization of the crystal
which can be accounted for by adjusting the
chemical potential~$\mu^{\pm}$ of the added or
removed electron accordingly by
using a continuum approximation.~\cite{Grafenstein:VB-93,%
Fulde:EC-95,Grafenstein:VB-97,Albrecht:AA-00}
This procedure is not specific to
the CO-ADC theory advocated here to
treat the short-range interactions, but is required
generally.

Having introduced a convenient method to truncate the configuration space
for CO-ADC, we have to elucidate the physical impact of configuration
selection on band structures.
To this end, we consider the fully translational
symmetry adapted form~(\ref{eq:M_k_hermitian}).
The modified coupling amplitudes~(\ref{eq:m_k_h_U}) in the band structure
matrix~(\ref{eq:solvepADC}) of a crystal carry two independent crystal
lattice vectors~$\vec g_1^{\,\prime}$ and $\vec g_2^{\,\prime}$.
If $\vec g_1^{\,\prime}$ and $\vec g_2^{\,\prime}$ are sufficiently far away
from the origin cell,
all two-electron matrix elements~(\ref{eq:m_R_h_U}) which contribute to
the modified coupling amplitude~$\bar U_{\sigma \; ; \;
\vec g_1^{\,\prime} \; \vec g_2^{\,\prime} \; \kappa^{\prime} \,
\alpha^{\prime} \, \beta^{\prime}}^{\pm}(\vec k)$
fall for all~$\sigma$ below the configuration selection
threshold and are therefore zero. Because the modified interaction
matrices~(\ref{eq:m_R_h_C}) vanish in the CO-ADC(2,2) approximation
and Eq.~(\ref{eq:m_k_h_KC}) thus denotes a diagonal matrix,
the eigenvalue associated with the configuration~$(\vec g_1^{\,\prime}
\; \vec g_2^{\,\prime} \; \kappa^{\prime} \, \alpha^{\prime} \, \beta^{\prime})$
is simply given by the diagonal element~$(\mat{\bar K}^{\pm}(\vec k))_{\vec g_1^{\,\prime} \;
\vec g_2^{\,\prime} \; \kappa^{\prime}
\, \alpha^{\prime} \, \beta^{\prime} \; ; \; \vec g_1^{\,\prime} \; \vec g_2^{\,\prime} \; \,
\kappa^{\prime} \, \alpha^{\prime} \, \beta^{\prime}}$,
with an eigenvector being unity on
position~$\vec g_1^{\,\prime} \; \vec g_2^{\,\prime} \; \kappa^{\prime}
\, \alpha^{\prime} \, \beta^{\prime}$ and
zero elsewhere. Consequently, such an eigenstate of~$\mat B(\vec k)$ is removed
by deleting the column and the row~$\vec g_1^{\,\prime} \; \vec g_2^{\,\prime} \;
\kappa^{\prime} \, \alpha^{\prime} \, \beta^{\prime}$
in~$\mat B(\vec k)$ because the transition
amplitude~(\ref{eq:transamp})
vanishes and thus it does not
contribute to the spectral representration of the one-particle Green's
function~(\ref{eq:spectral_CO-ADC}).

A further consequence of configuration selection involves
the lattice summations over~$\vec R$ in Eqs.~(\ref{eq:m_k_h_U})
and (\ref{eq:m_k_h_KC}). In the first place, we consider
a crystal with a macroscopic lattice constant which consists
of~$N_0$ unit cells.
As a result,
only Fock and two-electron matrix elements that involve
Wannier orbitals from
the origin cell are selected. Therefore,
in Eqs.~(\ref{eq:m_k_h_U}) and (\ref{eq:m_k_h_KC}),
the lattice sums run over only the single term
for~$\vec R = \vec 0$ with~$\vec g_1 = \vec g_2 = \vec 0$
and
the resulting band structure matrix~(\ref{eq:solvepADC}) is
consequently independent of~$\vec k$. As the Brillouin zone
contains $N_0$~crystal momenta,
we obtain a $N_0$-fold degenerate spectrum similarly to
simple tight-binding models.~\cite{Ashcroft:SSP-76}
In typical crystals, interactions with neighboring unit cells
are important. Yet a sufficiently large supercell which
consist of $n_0$~unit cells can be chosen such that interactions
between supercells are negligible leading to a $N_0 / n_0$~fold
degenerate spectrum.
In other words, a cutoff threshold, to be used for the selection of Fock
matrix elements and $2p1h$- and $2h1p$-configurations, implies a
definition of degeneracy of the physical states in our model
of the crystal.

In order to selectively diagonalize the band structure matrix, one uses
an iterative eigenvalue solver, \eg, a block-Lanczos
algorithm,~\cite{Lanczos:IM-50,Cullum:LA-85,Golub:MC-96,Weikert:BL-96}
that is capable of exploiting the sparsity of~$\mat B(\vec k)$.
The most expensive step of the block-Lanczos
algorithm is a matrix times vector product%
\footnote{Let $N_{\mathbf{A}}$~denote the number of
rows of the square matrix~$\mat A$. Then, specifically for the
CO-ADC(2) approximation in terms of Bloch orbitals~(\ref{eq:ADC-form-expanded}),
(\ref{eq:k-CO-ADC2}), and (\ref{eq:solvepADC}),
a single matrix times vector operation requires~$N\I{flops}
= 2 \, [N_{\mathbf F(\vec k)}^2 + (2 \, N_{\mathbf
  F(\vec k)} + 1) \, (N_{\mathbf K^+(\vec k)}
  + N_{\mathbf K^-(\vec k)}))]
= O(N_{\mathbf F(\vec k)} \,
  N_{\mathbf K^+(\vec k)})$ flops.
By affixing bar accents to the matrices showing up in
the expression for~$N\I{flops}$, we immediately obtain the
corresponding formula for the number
of flops required in the CO-ADC(2,2) approximation for
Wannier orbitals~(\ref{eq:M_k_hermitian}), (\ref{eq:R-CO-ADC2}),
and (\ref{eq:solvepADC}).}
between~$\mat B(\vec k)$ and
Lanczos vectors which determines the overall performance of the eigenvalue
solver.~\cite{Cullum:LA-85,Golub:MC-96,Weikert:BL-96,Buth:NH-04}
To investigate the scaling of a selective computation of eigenpairs,
we consider a supercell consisting of two unit cells of an
original lattice. Let us assume
a crystal with a macroscopic lattice constant to investigate
the asymptotic scaling behavior of the problem.
Then, the configuration selection method of the previous paragraphs
selects only configurations local to the individual unit cells.
Hence, the resulting configuration space
scales linearly with
the system size,
\ie, the number of atoms per unit cell.
This is a necessary condition for the
selective diagonalization of the band structure matrix~(\ref{eq:solvepADC})
to scale linearly as well.
The total Fock matrix of the supercell also is block-diagonal.
Similarly, the modified coupling amplitudes~(\ref{eq:m_k_h_U})
only contain two nonzero blocks which describe the coupling of
$2p1h$- and $2h1p$-configurations to the
Fock matrix of each of the two constituting unit cells of the
supercell such that the total band structure matrix decomposes
into two subproblems that can be solved independently.
A single matrix times vector product
for the supercell requires twice as many
floating point operations than are needed for a matrix
times vector product
for one of the two unit cells in the supercell. Hence, the
computation of matrix times vector products
scales linearly with the system size.%
\footnote{The number of (block-)Lanczos iterations,
necessary to determine the eigenvalues and eigenvectors of the band
structure matrix~(\ref{eq:solvepADC}) with a given accuracy, is assumed to be the same
for a single unit cell and a supercell consisting of two unit cells.}
Yet doubling the system size also usually means that we are interested
in twice as many excited states.
Hence, the overall
effort to determine all excited states of a crystal in a
given energy range
scales quadratically. It is a \emph{quadratically} scaling problem,
where linear scaling can only be achieved by an \emph{a priori}
restriction to a few excited states of the system.

For typical crystals, the Fock matrix of a supercell is not block-diagonal
which implies another doubling of the number of floating point operations
upon doubling the system size.
Yet this factor cancels as the number of $\vec k$-points needed for a
given accuracy of the integration over the Brillouin zone of the supercell
is halved because the volume of its
Brillouin zone is half the volume of the Brillouin zone
which corresponds to the original crystal
lattice.~\cite{Ashcroft:SSP-76}

\section{Lithium fluoride crystal}
\label{sec:LiF_crystal}

\begin{figure}
  \centering
  \includegraphics[width=5cm,clip]{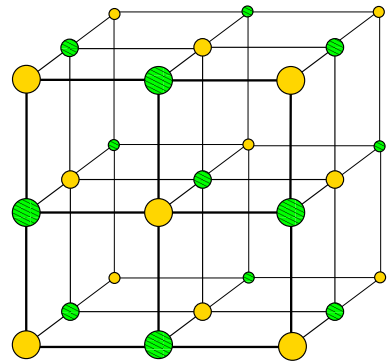}
  \caption{(Color) Structure of a LiF crystal. The lithium and
           fluorine atoms are represented by solid yellow and
           hatched green spheres, respectively.}
  \label{fig:lifgeom}
\end{figure}

Lithium fluoride crystallizes in a face-centered-cubic~(fcc)
rocksalt structure described by the space
group~$Fm\bar 3m$ as shown in Fig.~\ref{fig:lifgeom}.
The crystal lattice has a two-atomic basis
Li(0,0,0) and F($\frac{1}{2},
\frac{1}{2}, \frac{1}{2}$), given in units of
the lattice constant%
\footnote{This value for the lattice constant
of a LiF~crystal is utilized in
Refs.~\onlinecite{Prencipe:AI-95,Shukla:WF-98,Albrecht:LOB-02}
and is told to originate from the book of
Wyckoff.~\cite{Wyckoff:CS-68}
However, a value of~$4.0173 \angstrom$ is reported
therein which is consistent
with early and recent experimental values.
In order to allow comparison,
we adopt the previously used value of~$3.990 \angstrom$.
Although the deviation
is small, it has a quite noticeable influence on the
fundamental band gap
%
%
which decreased by~$0.65 \U{eV}$ when changing
from~$3.918 \angstrom$ to $4.026 \angstrom$ in
a recent density functional theory
study at the LDA~level.~\cite{Wang:QP-03}}
$a = 3.990 \angstrom$.
To investigate a LiF~crystal, we have chosen the simplest
possible \emph{ab initio} description.
The main purpose here is to demonstrate the
feasibility of the CO-ADC formulas for
a three-dimensional crystalline solid. Thus, only a minimal
Gaussian STO-6G basis set~\cite{Hehre:SC-69,basislib-04}
is used which
describes the lithium atom by a $(12s) \, / \, [2s]$~contraction
and the fluorine atom by~$(12s\,6p) \, / \, [2s\,1p]$.
Each shell
of the STO-6G basis set
is constructed by fitting six Gaussian
functions to each occupied orbital of the isolated
atoms.~\cite{Hehre:SC-69}
The $2s$~shell of a lithium atom is rather
diffuse due to the single outer valence electron.
As the lithium atoms are ionized
in LiF crystals, we remove the two most diffuse
Gaussian functions in the Li$\,2s$~contraction because
they
do not reflect the physical
situation of a compact Li$^+$~ion, and
arrive at a $(10s) \,/\, [2s]$~contraction
scheme.

The Hartree-Fock equations are solved self-consistently
by means of the \textsc{wannier}
program~\cite{Shukla:EC-96,Shukla:WF-98}
which directly yields Wannier orbitals.
A finite cluster of unit cells is utilized
as support for the Wannier orbitals of the origin cell
which, in our case, consist of
up to
third-nearest-neighbor cells (43~unit cells alltogether).
As the program determines
only occupied Wannier orbitals, crystal
projected atomic orbitals~(crystal PAOs)~\cite{Shukla:WFB-99,Buth:MC-05}%
$^,$\footnote{The threshold for the removal of redundancy in the
PAO construction was chosen such that the number of discarded
virtual Wannier orbitals per unit cell corresponds precisely
to the number of occupied Wannier orbitals per unit cell.~\cite{Buth:MC-05}}
have been devised
on the basis of the projected atomic orbitals introduced by Saeb\o{}
and Pulay~\cite{Pulay:LD-83,Saebo:LT-93} and also used by
Hampel and Werner.~\cite{Hampel:LT-96}
The resulting virtual functions are subsequently L\"owdin
orthonormalized.~\cite{Lowdin:QT-56,Lowdin:BT-62,Szabo:MQC-89,McWeeny:MQM-92}
For the CO-ADC(2,2) method~(\ref{eq:M_k_hermitian}), (\ref{eq:R-CO-ADC2}),
and (\ref{eq:solvepADC}), the Fock matrix and the
two-electron integrals in Wannier representation, entering the
Hamiltonian~(\ref{eq:Wannier_Hamilton}), need
to be obtained by a transformation of the corresponding
quantities
in terms of Gaussian basis
functions. This transformation is presently
the bottleneck in practical computations since the \textsc{wannier}
program~\cite{Shukla:EC-96,Shukla:WF-98}
carries it out for clusters only
and neither uses translational nor point group symmetry.
At present, this permits only the use of a minimal basis
set for a three-dimensional LiF~crystal. The subsequent
correlation calculations with the \textsc{co-adc}
program~\cite{co-adc-04,Buth:MC-05} are very efficient
and as rapid as typical molecular ADC calculations.

\begin{table}
  \centering
  \begin{ruledtabular}
    \begin{tabular}{ccccc}
      Cells & $E_{\rm top,v}$ & $E_{\rm bottom,c}$
      & $E_{\rm gap}$ & $\Delta E_{{\rm F}2p}$ \\
      \hline
      \phantom{0}0 & 0\phantom{.00} & 16.24 & 16.24 & 4.95 \\
      \phantom{0}1 & 0.41           & 16.16 & 15.74 & 4.80 \\
                13 & 2.43           & 15.69 & 13.26 & 4.78 \\
                19 & 2.89           & 15.36 & 12.47 & 4.75 \\
    \end{tabular}
  \end{ruledtabular}
  \caption{Convergence of
           the top of the valence
           bands~$E_{\rm top,v}$, the bottom of
           the conduction band~$E_{\rm bottom,c}$,
           the fundamental band gap~$E_{\rm gap}$,
           and the bandwidth of the F$\,2p$~valence
           band complex~$\Delta E_{{\rm F}2p}$ of a
           LiF~crystal.
           ``Cells''~designates the number of
           unit cells taken into account in
           CO-ADC(2,2) calculations of the
           quasiparticle band structure where zero
           refers to the Hartree-Fock result.
           Unity denotes the inclusion of the $2p1h$- and
           $2h1p$-configurations from the origin cell only
           in correlation calculations.
           13 and 19 indicate the additional inclusion of
           configurations from nearest- and next-nearest-neighbor cells,
           respectively.
           For a LiF crystal, $E_{\rm top,v}$ and
           $E_{\rm bottom,c}$ are both situated at the
           $\Gamma$~point.
           All data are given in electronvolts.}
  \label{tab:cabLiF}
\end{table}

\begin{figure}
  \centering
  \includegraphics[clip,width=\hsize]{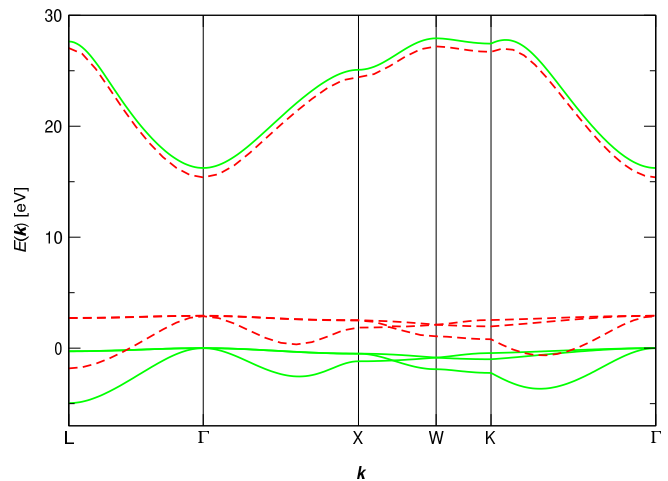}
  \caption{(Color) Band structure of a LiF crystal.
           Hartree-Fock bands are given by solid
           green lines. The dashed red lines depict the
           CO-ADC(2,2) quasiparticle bands which
           are determined by accounting
           for $2p1h$- and
           $2h1p$-configurations involving
           Wannier orbitals in the origin cell and the nearest- and
           next-nearest-neighbor cells.}
  \label{fig:LiF_19cell}
\end{figure}

\begin{figure}
  \centering
  \includegraphics[clip,width=\hsize]{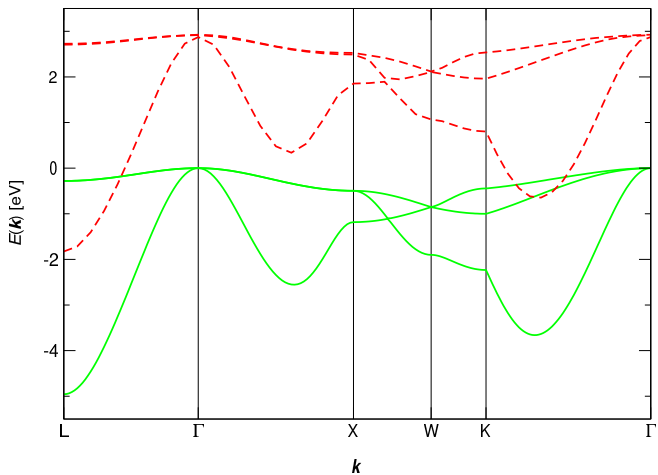}
  \caption{(Color) Valence band structure of a LiF~crystal.
           Zoom into Fig.~\ref{fig:LiF_19cell}.}
  \label{fig:LiF_19cell_valence}
\end{figure}

The Hartree-Fock band structure that results from this
minimal basis is shown in Fig.~\ref{fig:LiF_19cell}.
All band energies
are given with respect to the
top of the valence bands which is
located at the $\Gamma$~point
and is set to zero.
Below zero there is a complex
of three F$\,2p$~valence bands
shown as a closeup in
Fig.~\ref{fig:LiF_19cell_valence}. They originate
from the three F$\,2p$~valence energy
levels of the LiF molecule around $12 \U{eV}$
where the lowest ionization potential of the molecule
is doubly degenerate.~\cite{Buth:MC-05}
Upon crystallization,
this degeneracy is lifted
due to interactions with neighboring atoms
and, except for the high symmetry
lines~$\Gamma$--L and $\Gamma$--X,
one observes three distinct F$\,2p$~valence bands.
As a minimal basis set is utilized, there is
only a single conduction band
of mainly Li$\,2s$~character. A
slight admixture of F$\,2p$~character to this band
is observed in the
W--K~panel by comparing with the band structure resulting
from a larger basis set, \eg, the one of
Ref.~\onlinecite{Albrecht:LOB-02}, where a conduction
band of F$\,2p$~character mixes with the Li$\,2s$~band.
Nevertheless, the single
conduction band represents the energetically
lower edge of the conduction-band complex well.
LiF has a direct band gap;
\ie, the maximum of the F$\,2p$~valence bands and the
minimum of the Li$\,2s$~conduction band are located
at the same crystal momentum, the $\Gamma$~point, here.

In order to investigate electron correlations,
we start
from
a single unit cell,
to form the $2p1h$- and $2h1p$-configurations
entering Eq.~(\ref{eq:M_k_hermitian}) for~$\mat
M^{\pm}(\vec k,\omega)$, respectively.
Successively, configurations which contain Wannier orbitals in
nearest- and next-nearest-neighbor cells are considered.
This leads to three distinct quasiparticle band structures
whose convergence with respect to the number
of unit cells where electron correlations are
regarded explicitly is monitored for the following key
quantities: the top of the valence bands, the bottom of
the conduction band, and the band gap and the
width of the F$\,2p$~valence-band complex
[Tab.~\ref{tab:cabLiF}]. The results are
compared with the plain Hartree-Fock data
(with configurations out of ``0''~unit cells).
Taking into account electron correlations of a single
unit cell causes a slight
upwards shift of the
valence bands by~$0.41 \U{eV}$ while the conduction band
essentially remains unchanged.
Upon inclusion of configurations in the nearest-neighbor
cells, the band structure changes drastically,
leading to a significant reduction of the
band gap
%
%
%
by~$2.48 \U{eV}$. Accounting for configurations
which extend to next-nearest-neighbor cells has
%
%
three times less impact on the band gap.
%
%
%
Therewith, it can clearly be seen that the
major contributions of
electron correlations have been
covered
and that
the effect of configurations which extend to
Wannier orbitals
beyond next-nearest-neighbor cells will
yield much less significant changes of
the quantities than those discernible in Tab.~\ref{tab:cabLiF}.
As pointed out in Sec.~\ref{sec:onion}, the effect
of the more distant unit cells
can be determined using
a continuum approximation.~\cite{Grafenstein:VB-93,%
Fulde:EC-95,Grafenstein:VB-97,Albrecht:AA-00}
The observed correlation corrections
in Tab.~\ref{tab:cabLiF}
with a growing number
of neighboring unit cells result primarily from the
formation of a quasiparticle and the
polarization of the surrounding by the Coulomb charge of the added
particle or hole.~\cite{Fulde:EC-95}
Note that,
additionally, the beneficial basis set
extension~(BSE)~\cite{Pisani:HF-88,Duijneveldt:CP-94,Pisani:QM-96}
effect is noticeable in solids which leads to an
improved description of electron correlations
with an increasing number of unit cells taken into account.

The quasiparticle band structure of~LiF is displayed
in Fig.~\ref{fig:LiF_19cell} for the F$\,2p$~valence
bands and the Li$\,2s$~conduction band. The former bands are
additionally displayed on an enlarged scale
in Fig.~\ref{fig:LiF_19cell_valence}.
The data are taken from the
computation
with configurations from alltogether 19~unit cells.
Valence and conduction bands do not shift
by the same amount, upwards and downwards, respectively.
Instead, the former bands are much more strongly influenced by
electron correlations than the latter bands.
This is due to the fact that the
occupied Wannier orbitals are spatially less extended
than the virtual Wannier orbitals.
For electrons in compact orbitals, the Hartree-Fock
approximation becomes progressively worse
due to an insufficient description of the
Coulomb hole.~\cite{Fulde:EC-95}
Although the quasiparticle bands are considerably shifted
with respect to the corresponding Hartree-Fock bands,
they essentially keep their width.
We also observe this effect for~\HF~chains.~\cite{Buth:MC-05}
This is in contrast to the significant reduction of bandwidths
observed for quasiparticle band structures
of covalently bonded polymers
like \textit{trans}-polyacetylene~\cite{Bezugly:MC-04} and
covalently bonded crystals like diamond,~\cite{Grafenstein:VB-93,%
Grafenstein:VB-97,Albrecht:AA-00}
silicon~\cite{Grafenstein:VB-97,Albrecht:AA-00},
and germanium.~\cite{Grafenstein:VB-97}

The \emph{ab initio} description of a LiF~crystal
using a minimal basis set provides valuable
insights.
However, we cannot expect
full quantitative agreement neither at the Hartree-Fock
nor at the correlation level.
The accuracy of our approach can be determined by
comparing with experimental and theoretical data
from related studies.
The fundamental band gap is accessible by
photoelectron spectroscopy and
has been measured to be~$14.1 \U{eV}$.
Another experimentally accessible quantity, the
width of the F$\,2p$~valence-band complex, was
determined to lie in the range~\cite{Kunz:SE-82}
of~$3.5$--$6 \U{eV}$
and thus is inconclusive, unfortunately.
With our STO-6G-like basis set, we find
a Hartree-Fock band gap of~$16.24 \U{eV}$ [Tab.~\ref{tab:cabLiF}]
which deviates considerably from
the value of~$22.7 \U{eV}$ communicated by Kunz~\cite{Kunz:SE-82}
or the value of~$22.4 \U{eV}$ reported by Albrecht~\cite{Albrecht:LOB-02}
who both use larger basis sets.
For the width of the F$\,2p$~valence-band complex,
we find~$4.95 \U{eV}$ [Tab.~\ref{tab:cabLiF}] in the Hartree-Fock approximation
which is appreciably larger than
the value of~$3 \U{eV}$ from Kunz~\cite{Kunz:SE-82} and
the value of~$3.37 \U{eV}$ from Albrecht.~\cite{Albrecht:LOB-02}

Including electron correlations, the quasiparticle
band gap reduces to~$12.47 \eV$ [Tab.~\ref{tab:cabLiF}] and thus
falls short by~$1.5 \eV$ compared to the experimental value.
Albrecht~\cite{Albrecht:LOB-02} finds $13.5 \U{eV}$,
and calculations of Kunz~\cite{Kunz:SE-82}
yield a theoretical band gap of~$14.0 \U{eV}$,
in good agreement with the recent measurements of~$14.1 \U{eV}$.
For the F$\,2p$~valence band complex, we obtain a width of~$4.75 \U{eV}$
[Tab.~\ref{tab:cabLiF}]
which is nearly unchanged compared to the
Hartree-Fock value,
a fact that has also been found in previous studies
of bulk~LiF.~\cite{Kunz:SE-82,Albrecht:LOB-02}
Consequently, our value of the width remains
much higher than the
value of~$3.1 \U{eV}$ from Kunz~\cite{Kunz:SE-82}
or the
value of~$3.40 \U{eV}$ from Albrecht.~\cite{Albrecht:LOB-02}

The above results show that the CO-ADC theory can be
applied without problems to compute the energy
bands of three-dimensional crystalline solids.
What remains to be improved considerably
is the transformation of the Fock matrix and
the two-electron integrals from the Gaussian
basis set representation to a representation
in terms of Wannier orbitals.

\section{Conclusion}
\label{sec:conclusion}

In this article, a general local orbital \emph{ab initio} Green's
function method is devised for band structures. The
band structure is given by the pole positions of the
one-particle Green's function in terms of
Bloch orbitals which in turn is
evaluated efficiently utilizing Dyson's equation.
To approximate the self-energy up to $n$-th order,
we devise a crystal orbital formulation of the well-established algebraic
diagrammatic construction~(ADC)~\cite{Schirmer:PP-82,%
Schirmer:GF-83,Cederbaum:GF-98} which is
termed crystal orbital ADC~(CO-ADC).~\cite{Deleuze:SC-03} The
pole search of the one-particle Green's function is recast
into a Hermitian eigenvalue problem which is a numerically stable and
efficient formulation, permitting us to explore strong
correlations that occur for the energetically lower
lying bands of crystals.~\cite{Lowdin:QT-56,Lowdin:BT-62,%
Ashcroft:SSP-76,Cederbaum:CE-86,Buth:IO-03,Buth:IM-03}
Wannier orbitals are used in
internal summations in the self-energy and
allow us to exploit the fact
that electron correlations are predominantly local.

The lattice summations in the CO-ADC equations
must be truncated to render the
problem tractable. To this end, we devise a configuration selection
procedure which can equally well be used in conjunction
with the calculation of ionization potentials and electron
affinities of large molecules to speed up computations.
Configuration selection is shown to lead to a definition of
degeneracy among the states of a crystal.

The derivation of the local orbital CO-ADC theory
sets out
from
the equations in terms of Bloch orbitals. Afterwards internal indices are
transformed to a Wannier representation. We consider this line of
argument to be compelling due to the close analogy of the
equations in crystal momentum representation to the equations
of molecular physics.
Alternatively, the derivation can be conducted by starting from CO-ADC in terms
of Wannier orbitals. A transformation of the equations to
crystal momentum representation is then carried out by utilizing the
Wannier transformation~(\ref{eq:Wannier}).

The \textsc{co-adc} computer program~\cite{co-adc-04,Buth:MC-05}
has been developed for computations based on the CO-ADC(2,2) approximation.
It requires the Fock matrix and
the two-electron integrals in local representation
which are obtained by an integral transformation using
the Wannier orbitals from a Hartree-Fock calculation
with the \emph{ab initio} program
\textsc{wannier}.~\cite{Shukla:EC-96,Shukla:WF-98}
At present, this transformation poses a bottleneck
because the current implementation has been designed
for molecular clusters and does not exploit any symmetries.
Therefore, only a minimal basis set is used
to study a lithium fluoride crystal
in order to demonstrate that the method is working.
A fundamental band gap of~$12.47 \U{eV}$ and a bandwidth
of the F$\,2p$~valence-band complex of~$4.75 \U{eV}$ is obtained.
These values are to compare with the experimental data
of~$14.1 \U{eV}$ and $3.5$--$6 \U{eV}$ and the theoretical
results of~$13.5$, $14.0 \U{eV}$, and $3.1$, $3.4 \U{eV}$
for the band gap and the bandwidth, respectively.
Increasing the size of the basis set
will improve the CO-ADC(2,2) results.
Yet it is demonstrated that the CO-ADC theory
is well suited for tackling the
problem of calculating energy bands of solids
with controlled approximations.

We would like to point out that the block of $2p1h$-configurations in the
band structure matrix~(\ref{eq:solvepADC}) is much larger than the block of
$2h1p$-configurations. In order to reduce the computational effort, the advanced part and
the retarded part of the one-particle Green's function can be
independently evaluated diagrammatically.
This leads to a different, so-called non-Dyson CO-ADC scheme,
which decouples the
computation of occupied and virtual bands completely~\cite{Schirmer:ND-98,%
Birkenheuer:ND-04} and results in two independent Hermitian matrix
eigenvalue problems to be solved.

\begin{acknowledgments}
We would like to thank Jochen Schirmer for fruitful discussions
and critically reading the manuscript.
C.B.~is indebted to Thomas Sommerfeld for providing his
complex-symmetric block-Lanczos algorithm and helpful advice.
\end{acknowledgments}

\end{document}